\documentclass[aps,preprint,nofootinbib,nobibnotes,natbib]{revtex4}

\usepackage{orcidlink}
\usepackage{amsmath}
\usepackage{amssymb}
\usepackage{graphicx}
\usepackage{epstopdf}
\usepackage{color}
\usepackage[normalem]{ulem}
\usepackage{soul}
\usepackage[FIGTOPCAP]{subfigure}
\usepackage{tabularx}
\usepackage{enumitem}
\newenvironment{mybibliography}[1]{
    \begin{enumerate}[label={[\arabic*]}, ref=\arabic*, leftmargin=*]
    \setcounter{enumi}{\value{enumiv}} 
}{
    \end{enumerate}
}

\usepackage[mathlines]{lineno}

\def\aap{Astron.\ Astrophys.\ }

\renewcommand{\thefootnote}{\fnsymbol{footnote}}

\begin{document}

\title{Charge-dependent spectral softenings of primary cosmic-rays below the knee}


\author{
DAMPE Collaboration$^{\ast}$
}

\footnotetext[1]{Members of DAMPE Collaboration are listed at the end of this paper. \\Email: dampe@pmo.ac.cn}

\begin{abstract}
In most particle acceleration or propagation theories, the characteristic features of the 
cosmic ray spectra due to acceleration limits or propagation phase changes are charge
dependent\cite{Peters:1961mxb,Hillas:1984ijl,Bell:2013vxa,Cesarsky:1980pm}. Alternatively,
the interaction scenario would expect mass dependent spectral features in general. 
The observational verification of which relation takes effect in nature is still lack due 
to the difficulty of measuring the spectra of individual particles up to very high energies. 
Here we report direct measurements of the carbon, oxygen, and iron spectra from $\sim 20$ 
gigavolts to $\sim 100$ teravolts ($\sim 60$ teravolts for iron) with 9 years of on-orbit 
data collected by the Dark Matter Particle Explorer. Distinct spectral softenings 
have been directly detected in these spectra for the first time. Combined with the updated 
proton and helium spectra, the spectral softening appears universally at a rigidity of $\sim15$ 
teravolts. A nuclei mass dependent softening is rejected at a confidence level of $>99.999$\%. 
Possible interpretations of these results, including a nearby cosmic ray 
source\cite{Savchenko:2015dha,Ahlers:2016njd,Liu:2018fjy} and other models such as the
propagation effect\cite{Chernyshov:2022kxk}, are discussed. 
\end{abstract}

\maketitle
\normalsize

Cosmic rays (CRs) are energetic particles travelling through the Universe as high-energy beams.
They are predominantly made up of nuclei, such as protons, heavier ions up to iron and 
beyond\cite{Hillas:1984ijl}. Depending on the production processes, CRs are typically divided 
into two classes, the primary species accelerated by sources and secondary species produced due 
to fragmentation of primary CRs via collisions with the interstellar medium (ISM). High abundance 
particles such as protons, helium, carbon, oxygen, and iron nuclei are believed to be dominated 
by primary origin, while lithium, beryllium, boron, and sub-iron nuclei belong to the secondary 
family\cite{Strong:2007nh}. 
Spectral structures of CRs are very important probes to understanding the physics of CRs such 
as the acceleration, propagation, and interactions. In the general picture of particle acceleration 
and propagation, charge-dependent spectral features are expected due to gyration of particles
in the magnetic fields. On the other hand, in case of particle interactions with characteristic 
thresholds, the break off energy of particles would be mass dependent.
Obtaining observational evidence to verify which factor shapes the CR spectra is highly challenging 
and remains elusive. To convincingly test this, precise spectra of individual CR species covering a 
wide range of charge numbers need to be measured up to very high energies, which requires sensitive
detectors with large acceptance and good charge/energy resolution.    

Important progresses on precise measurements of energy spectra of different mass compositions
have been achieved in recent years. Direct measurements of the spectra of individual CR species by
space-borne and balloon-borne experiments have revealed unexpected hardening features around a few
hundred GV rigidity\cite{PAMELA:2011mvy,Panov:2009iih,AMS:2017seo,AMS:2021nhj,DAMPE:2019gys,Alemanno:2021gpb,DAMPE:2022jgy,DAMPE:2024qwc,CALET:2022vro,CALET:2023nif}, deviating from the power-law (PL) 
expectation from the conventional theories of the acceleration and propagation of CRs. In addition, 
spectral softenings around $\sim 10-30$ TeV have also been revealed in the proton and helium
spectra\cite{Atkin:2018wsp,DAMPE:2019gys,Alemanno:2021gpb,CALET:2022vro,CALET:2023nif,Choi:2022aht} 
(there were also hints from earlier experiments\cite{Panov:2009iih,Yoon:2017qjx,Gorbunov:2018stf}, 
but with relatively large statistical and systematic uncertainties). These new findings challenge 
the standard CR paradigm that a single source population with PL spectra accounts for the measured CR 
spectra below the so-called knee\cite{KK1958}, giving new insights on the understanding of CR physics. 
Though also being among the most abundant species in CRs, the measurements of carbon, oxygen, and 
iron nuclei spectra in CRs are rather challenging, and the current precise measurements are limited 
to a rigidity of a few TV. Similar hardening features as lighter nuclei have been detected for carbon 
and oxygen nuclei\cite{AMS:2017seo,Adriani:2020wyg}, while the iron spectrum is found to be consistent 
with a single PL above 100 GV\cite{AMS:2021lxc,CALET:2021fks}. The high-energy spectral behaviors of 
these medium and heavy nuclei are unclear yet, which hinders a critical understanding of the 
acceleration and propagation physics of CRs.

In this work, we report the direct measurements of the carbon, oxygen, and iron spectra up to
PeV energies with the Dark Matter Particle Explorer (DAMPE)\cite{DAMPE:2017cev,DAMPE:2017fbg}. 
DAMPE is a space-borne, calorimeter-type detector for high-energy particles and photons. With the 
plastic scintillator detector (PSD\cite{Yu:2017dpa}), the silicon tungsten tracker-converter 
(STK\cite{Azzarello:2016trx}), the bismuth germanium oxide calorimeter (BGO\cite{Zhang:2016xkz}), 
and the neutron detector (NUD\cite{Huang:2020skz}), DAMPE is able to measure precisely the charge,
direction, energy and particle identity of high-energy CRs. Particularly, the thick calorimeter (with $\sim32$ radiation lengths) is designed for observing CR electrons/positrons and $\gamma$ rays 
to energies beyond 10 TeV with a very high energy resolution\cite{DAMPE:2017fbg}. The relatively 
large geometric factor ($\sim0.3$ m$^2$~sr) and nuclear interaction depth ($\sim1.6\lambda_0$) make 
it also a powerful detector of CR nuclei to hundreds of TeV energies\cite{DAMPE:2019gys,Alemanno:2021gpb}.

\begin{figure}[!htbp]
\begin{center}
\includegraphics[width=0.7\columnwidth]{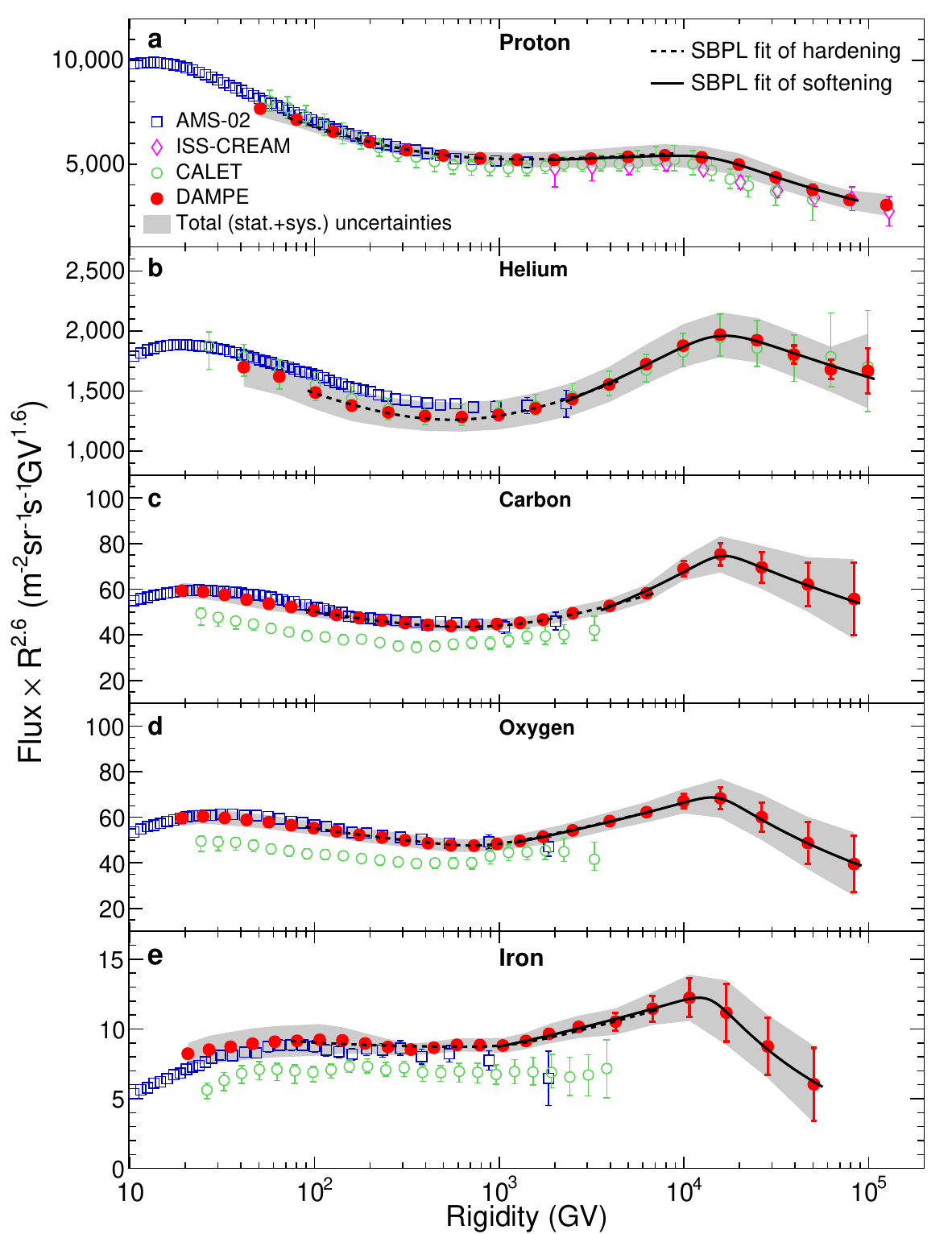}
\end{center}
\caption{ {\bf Rigidity spectra.} Panels {\bf a} - {\bf e} show the spectra of protons, helium, carbon, oxygen, and iron, 
weighted by $R^{2.6}$. Red dots are the DAMPE measurements, with $\pm1\sigma$
statistical uncertainties (errorbars) and the total statistical and systematic
uncertainties added in quadrature (shaded bands). 
Dashed and solid lines show the best-fitting results of the SBPL function.
The fluxes are converted from the total kinetic energy bins to rigidity bins assuming 
a mass number of 12, 16, and 56 for carbon, oxygen, and iron. For proton and helium, 
the average mass number is calculated assuming D/p$=0.027$ and $^{3}$He/$^{4}$He$=0.142 
\times (R/4.5~\text{GV})^{-0.289}$ based on the AMS-02 measurements\cite{AMS:2024idr}. 
Also shown are the measurements from other experiments: 
AMS-02\cite{AMS:2017seo,AMS:2021lxc,AMS:2021nhj}, 
CALET\cite{Adriani:2020wyg,CALET:2021fks,CALET:2022vro,CALET:2023nif}, and the Cosmic Ray 
Energetics And Mass on the International Space Station (ISS-CREAM)\cite{Choi:2022aht}.
}
\label{fig-1}
\end{figure}

The analysis presented in this work is based on the data recorded in the first 9 years of 
DAMPE's on-orbit operation, from January 1, 2016 to December 31, 2024. The live time fraction 
is about 76.2\% after excluding the instrument dead time, the time for on-orbit calibration, 
the time in the South Atlantic Anomaly (SAA) region, and the period between September 9, 2017 
and September 13, 2017 during which a big solar flare occurred and affected the status of the
detector\cite{DAMPE:2021qet}. The carbon, oxygen, and iron nuclei are efficiently identified 
based on the PSD charge measurements. The residual backgrounds for selected candidates are 
evaluated based on the template fitting for the Monte Carlo (MC) simulated charge distributions 
(see the Methods for details). The selection efficiency and the energy response are also 
obtained with MC simulations, and are validated based on the flight data and the test beam data.
An unfolding procedure\cite{DAgostini:1994fjx} is applied to the observed count spectrum 
to correct the bin-by-bin migration of events due to the limited energy resolution of the detector.  
The obtained energy spectra of carbon, oxygen, and iron nuclei, together with the updated proton 
and helium spectra, in the rigidity range from $\sim20$ GV to $\sim100$ TV (60 TV for iron) are 
shown in Figure 1. Compared with previous measurements of carbon, oxygen, and iron 
nuclei by the Alpha Magnetic Spectrometer (AMS-02\cite{AMS:2017seo,AMS:2021lxc}) and
the Calorimetric Electron Telescope (CALET\cite{Adriani:2020wyg,CALET:2021fks}), the DAMPE 
results improve the precision considerably at high energies and provide the first precise 
measurements of these spectra above several TV.

It can be clearly found that similar spectral structures, i.e. a hardening at $\sim500-1000$ GV 
rigidity followed by a softening at $\sim15$ TV, are observed among all the five species of CRs. 
The spectral fitting with a smoothly broken power-law (SBPL) model (see the Methods) provides 
significance of the hardening of $29\sigma$, $23\sigma$, $11\sigma$, $10\sigma$, and $2.7\sigma$ 
with hardening rigidity at $590\pm40$, $595\pm40$, $892\pm210$, $799\pm76$, and $1104\pm435$ GV 
for proton, helium, carbon, oxygen, and iron, respectively. The smoothness parameter of
the break can also be obtained (except for iron), as given in Table \ref{Table:fit_para_hard}. 
The results show that in general a smooth hardening is favored by 
the data. Although the spectra of all these CR species show similar hardening features, their 
detailed spectral shapes and break rigidities (energies) differ from each other. 
Fitting to the AMS-02\cite{AMS:2021nhj} and CALET measurements give also diverse results 
of the hardening rigidities\cite{Niu:2022vpn,CALET:2022vro,CALET:2023nif,Adriani:2020wyg}.
Measurements of the secondary-to-primary ratios of CRs such as B/C and B/O show hardenings 
around 200 GV\cite{AMS:2018tbl,AMS:2021nhj,DAMPE:2022jgy}, which is different from the break 
rigidities found here for primary CRs. This may suggest that the hardenings of primary and 
secondary CRs are not solely due to the same origin such as a unified propagation
effect\cite{Blasi:2012yr}.

What is more interesting is the spectral softening which is for the first time clearly identified in the
spectra of carbon, oxygen, and iron. The significance of the softening is 3.2$\sigma$, 4.1$\sigma$ and
2.4$\sigma$ with the break rigidity of $16.1\pm5.6$, $15.4\pm4.8$, and $13.8\pm6.0$ TV for carbon, oxygen, 
and iron, respectively. The updated proton and helium spectra yield softening rigidity at $14.6\pm1.3$ 
and $15.3\pm2.3$ TV, respectively. It has been clearly established that the softenings occur at the same
rigidity for all the five species, as shown in Figure 2. The proportional relation between the
particle charge $Z$ (for the four species with $Z\ge2$) and the softening energy $E_{\rm br}$ is best
fitted as $\bar{E}_{\rm br}/{\rm TeV} = (15.3\pm 1.6) \times Z$, which is well consistent with the 
$E_{\rm br}$ of proton, i.e. $14.6 \pm 1.3$ TeV. Assuming a general power-law scaling of the 
break energy and the charge, we get the fitting power-law index of $1.01\pm0.09$, which shows a good
consistency with the linear correlation. On the other hand, the mass number ($A$) dependence of the
softening energies can be excluded at a confidence level of $> 99.999$\% (4.4$\sigma$) by comparing 
the $E_{\rm br}/{\rm TeV} = 14.6 \pm 1.3$ for proton and the $\bar{E}_{\rm br}/{\rm TeV} = (7.6 \pm 0.9) 
\times A$ for other four species with $Z\ge2$. The universal softening feature in the spectra of primary 
CRs from protons ($Z=1$) to iron nuclei ($Z=26$) and the charge-dependence of the softening energy 
indicate a common origin of the break due to acceleration or propagation effect.

\begin{figure}[!htbp]
\begin{center}
\includegraphics[width=0.49\columnwidth]{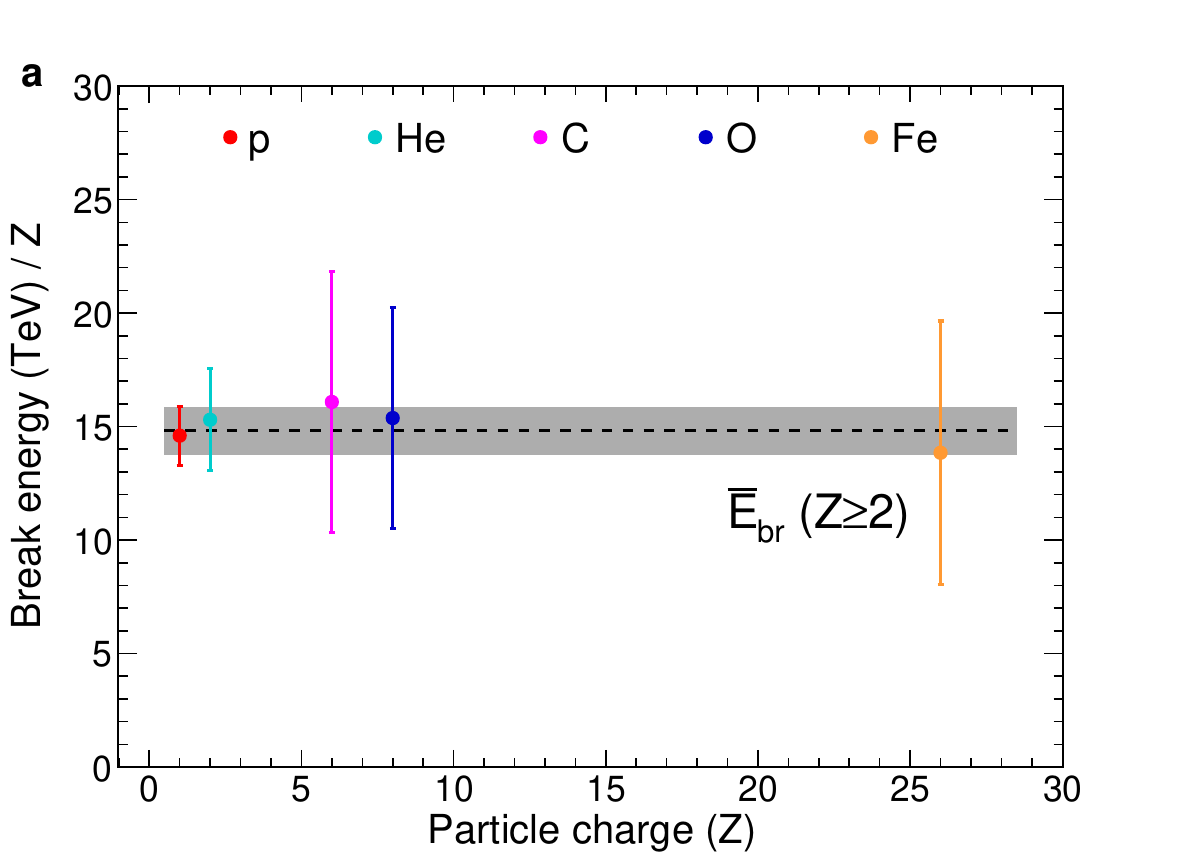}
\includegraphics[width=0.49\columnwidth]{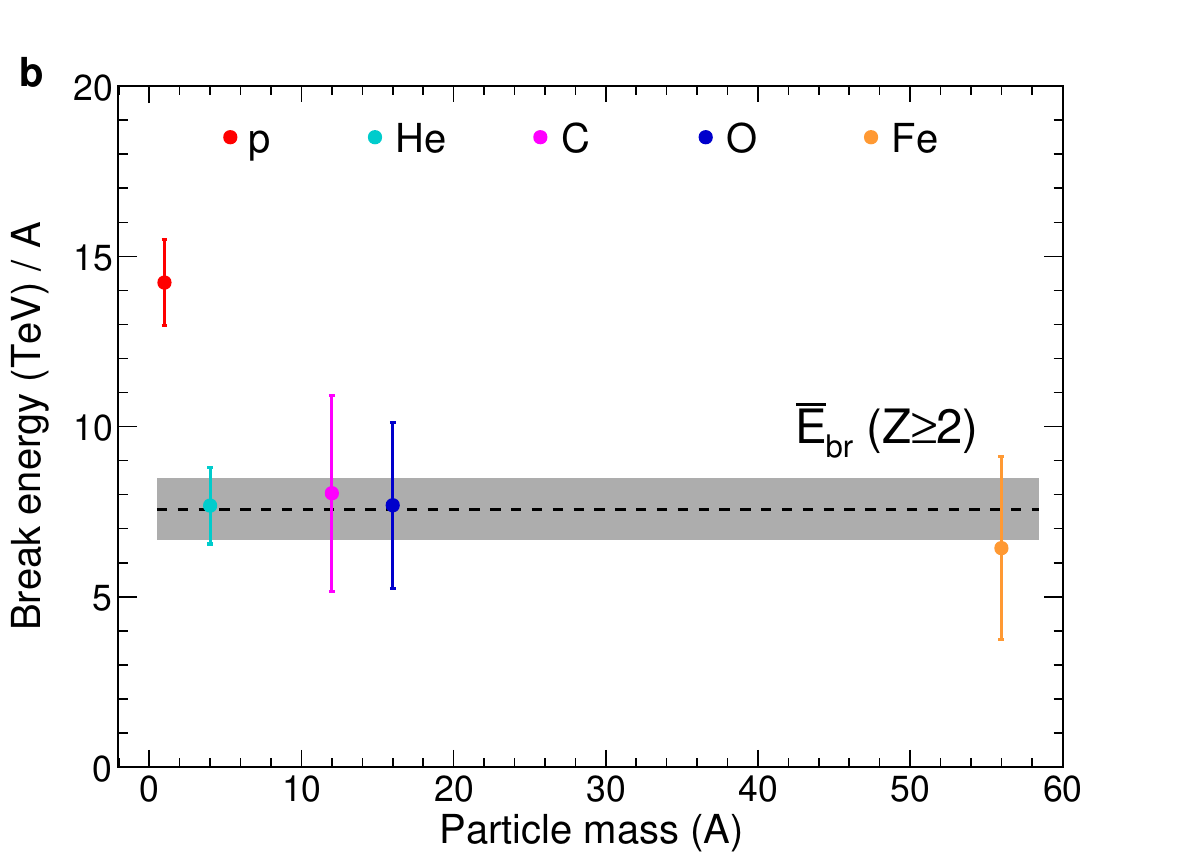}
\end{center}
\caption{
{\bf The break energy for different species.}
Panel {\bf a} shows the break energy divided by $Z$ as a function of particle charge $Z$, 
and panel {\bf b} is that divided by $A$ as a function of mass number $A$. 
The average mass number $A$ is assumed to be 1.026, 3.985, 12, 16, and 56 for proton, helium, 
carbon, oxygen, and iron, respectively. The shaded bands represent the average softening 
energies of the particles heavier than proton.
}
\label{fig-2}
\end{figure}

The spectra can be decomposed into two components, a soft component which dominates 
the CR fluxes in the low-energy band, and a hard component with a spectral cutoff 
which contributes mainly to the bump feature. A simple idea to account for the spectral 
structures is the superposition of a nearby source on top of the background source
population\cite{Savchenko:2015dha,Ahlers:2016njd,Liu:2018fjy}, as shown by solid lines
in Figure 3. The addition of the nearby source component with proper distance 
and age can simultanesouly reproduce the observed energy evolution of the amplitudes 
and phases of the dipole anisotropies (see Figure \ref{fig-aniso}).
See the Methods for details of the model setting. 
In this scenario, the spectrum from the nearby source shows both low-energy suppression 
due to the inefficient propagation of these particles within the age of the source and 
high-energy suppression due to the acceleration limit, and hence produces naturally a 
bump structure. Assuming an exponential form of the spectral cutoff, we obtain a cutoff 
rigidity of about 30 TV for the nearby source, which is consistent with the estimation 
of acceleration by supernova remnants\cite{Lagage:1983zz}. 
Supposing the spectral features are indeed imprints from a nearby source, we propose that 
the supernova explosion associated with the Geminga pulsar could be a good candidate for this 
source, given its proper age, distance, and location in the sky\cite{Liu:2018fjy,Zhao:2021css}. 
The energetics needed to account for the data is about $3.3\times 10^{50}$ erg, which is 
consistent with that can be produced for a typical core collapse supernova. 
Via comparing the relative abundances of nuclei at $R=0.1$ TV and $R=10$ TV, we can 
infer that the metallicity of the nearby source may be slightly higher than the average 
value of the background source population, as shown in Figure \ref{fig-abun}.

\begin{figure}[!htbp]
\begin{center}
\includegraphics[width=0.49\columnwidth]{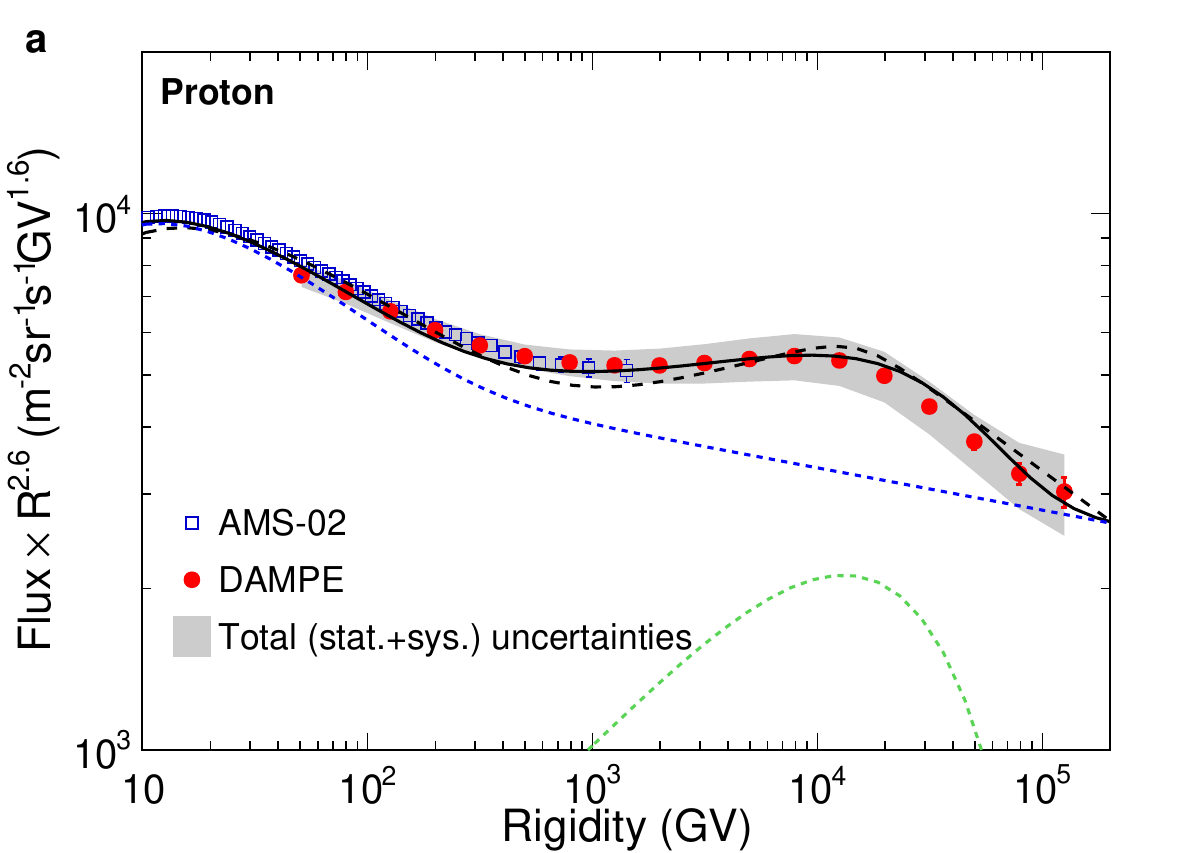}
\includegraphics[width=0.49\columnwidth]{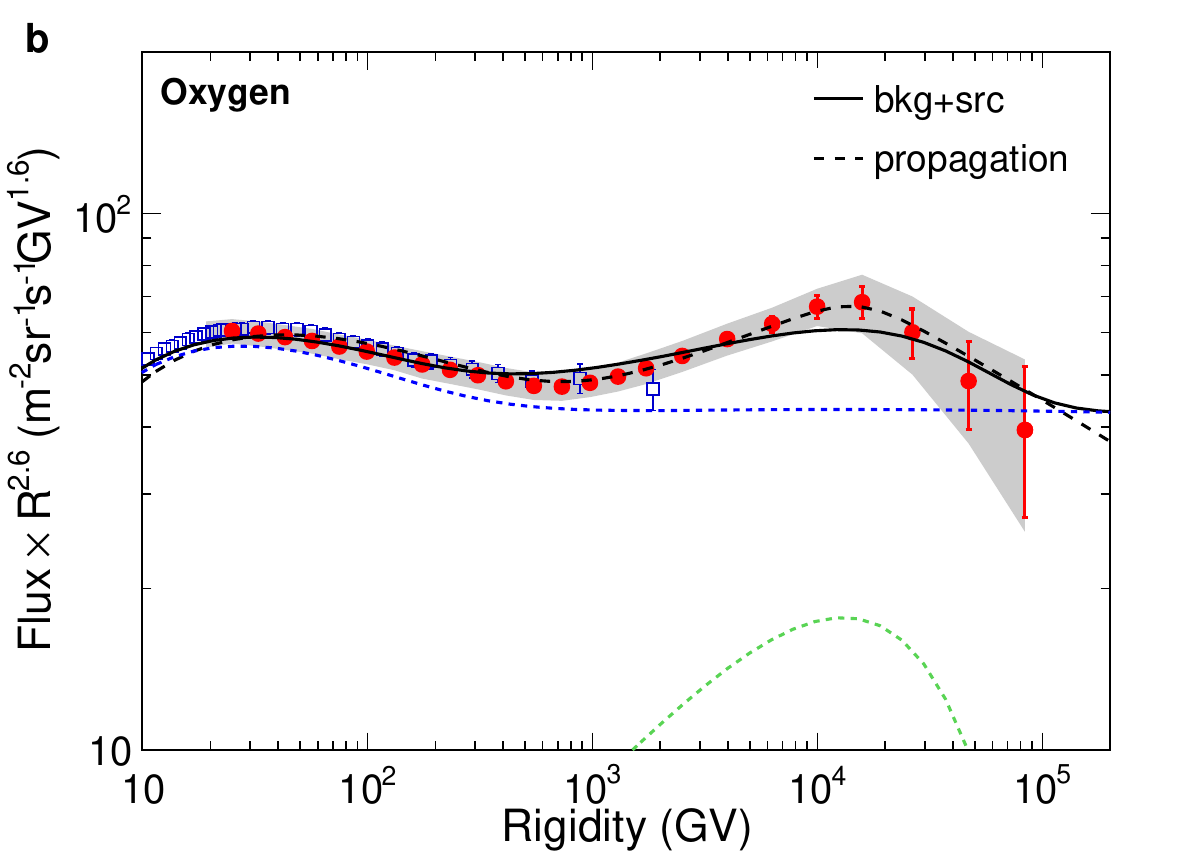}
\end{center}
\caption{
{\bf Comparison of the spectra with model expectations.}
Panel {\bf a} is for protons, and panel {\bf b} is for oxygen nuclei.
Solid lines show the model expectations of the background plus nearby source scenario, 
with individual contributions from the two components being shown by dotted lines 
(blue for the background and green for the nearby source), 
and dashed lines show the results of the propagation scenario. 
}
\label{fig-3}
\end{figure}

Alternatively, the spectral breaks could be due to a propagation effect, especially
that universal spectral hardenings have been found in the secondary-to-primary 
ratios\cite{AMS:2018tbl,AMS:2021nhj,DAMPE:2022jgy}. The physical modeling of the effect of 
self-generated turbulence with nonlinear Landau and ion-neutral damping of magnetohydrodynamic 
waves by CRs can properly account for both the hardening and softening of the proton
spectrum\cite{Chernyshov:2022kxk}. As an effective approach, we introduce two breaks 
of the diffusion coefficient to mimic the change of propagation properties, and calculate
the propagated spectra. The results are shown by dashed lines in Figure 3. 
The detailed parameter setting can be found in the Methods. The spectral structures 
observed by DAMPE and AMS-02 can also be properly reproduced by this model.

Additional theoretical models to account for the bump structures include multiple source
populations\cite{Bowman:2022okd,Recchia:2023mmg}, and the re-acceleration of CRs by a local 
star\cite{Malkov:2021gxd}. All these models can properly explain the features of the spectra. 
To explain the large-scale anisotropies in these scenarios, additional assumptions about 
the source distribution and/or magnetic field configuration are necessary\cite{Mertsch:2014cua}.
The DAMPE results reported in this work demonstrate a clear presence of rigidity-dependent
softening features in the spectra of primary CR nuclei, indicating a common origin due to
acceleration and/or propagation processes.

\bigskip
{\bf Acknowledgments:} The DAMPE mission was funded by the strategic priority science and technology projects in space science of Chinese Academy of Sciences (CAS). In China, the data analysis is mainly supported by the National Natural Science Foundation of China (No. 12588101) and the National Key Research and Development Program of China (No. 2022YFF0503302). Additional supports are from the CAS Project for Young Scientists in Basic Research (Nos. YSBR-061 and YSBR-092), the NSFC (Nos. 12220101003, 12503107), the Strategic Priority Program on Space Science of CAS (No. E02212A02S), the Youth Innovation Promotion Association of CAS, the Young Elite Scientists Sponsorship Program by CAST (No. YESS20220197), the Natural Science Foundation of Jiangsu Province (No. SBK20250300783), and the Program for Innovative Talents and Entrepreneur in Jiangsu. Y.Z.F thanks the support of New Cornerstone Science Foundation through the XPLORER PRIZE. In Europe, the activities and data analysis are supported by the Swiss National Science Foundation (SNSF), Switzerland, the National Institute for Nuclear Physics (INFN), Italy, and the European Research Council (ERC) under the European Union's Horizon 2020 research and innovation programme (No. 851103), and the Swiss State Secretariat for Education, Research and Innovation (SERI).

\renewcommand{\refname}{References}

\clearpage
\renewcommand{\thefootnote}{\fnsymbol{footnote}}
\centerline
{\large\bf DAMPE collaboration}
\begin
{flushleft}
\noindent
\small
Francesca~Alemanno$^{1,2}$\orcidlink{0000-0003-1065-2590},
Qi~An$^{3,4}$\footnote{Deceased.},
Philipp~Azzarello$^{5}$,
Felicia-Carla-Tiziana~Barbato$^{6,7}$\orcidlink{0000-0003-0751-6731},
Paolo~Bernardini$^{1,2}$\orcidlink{0000-0002-6530-3227},
Xiao-Jun~Bi$^{8,9}$,
Hugo~Boutin$^{5}$\orcidlink{0009-0004-6010-9486},
Irene~Cagnoli$^{6,7}$\orcidlink{0000-0001-8822-5914},
Ming-Sheng~Cai$^{10,11}$\orcidlink{0000-0002-9940-3146},
Elisabetta~Casilli$^{6,7}$\orcidlink{0009-0003-6044-3428}
% ...
\bigskip

{\footnotesize \it
$^1$Dipartimento di Matematica e Fisica E. De Giorgi, Universit\`a del Salento, I-73100, Lecce, Italy\\
$^2$Istituto Nazionale di Fisica Nucleare (INFN) - Sezione di Lecce, I-73100, Lecce, Italy\\
$^3$State Key Laboratory of Particle Detection and Electronics, University of Science and Technology of China, Hefei 230026, China\\
$^4$Department of Modern Physics, University of Science and Technology of China, Hefei 230026, China\\
$^5$Department of Nuclear and Particle Physics, University of Geneva, CH-1211, Switzerland\\
$^6$Gran Sasso Science Institute (GSSI), Via Iacobucci 2, I-67100 L’Aquila, Italy\\
$^7$Istituto Nazionale di Fisica Nucleare (INFN) - Laboratori Nazionali del Gran Sasso, I-67100 Assergi, L’Aquila, Italy\\
$^8$University of Chinese Academy of Sciences, Beijing 100049, China\\
$^9$Particle Astrophysics Division, Institute of High Energy Physics, Chinese Academy of Sciences, Beijing 100049, China\\
$^{10}$Key Laboratory of Dark Matter and Space Astronomy, Purple Mountain Observatory, Chinese Academy of Sciences, Nanjing 210023, China\\
$^{11}$School of Astronomy and Space Science, University of Science and Technology of China, Hefei 230026, China\\
$^{12}$State Key Laboratory of Heavy Ion Science and Technology, Institute of Modern Physics, Chinese Academy of Sciences, Lanzhou 730000, China\\
$^{13}$National Space Science Center, Chinese Academy of Sciences, Nanertiao 1, Zhongguancun, Haidian district, Beijing 100190, China\\
$^{14}$Istituto Nazionale di Fisica Nucleare, Sezione di Bari, via Orabona 4, I-70126 Bari, Italy\\
$^{15}$Dipartimento di Fisica ``M.~Merlin'', dell’Universit\`a e del Politecnico di Bari, via Amendola 173, I-70126 Bari, Italy\\
$^{16}$Istituto Nazionale di Fisica Nucleare (INFN) - Sezione di Perugia, I-06123 Perugia, Italy\\
$^{17}$Department of Physics and Laboratory for Space Research, the University of Hong Kong, Hong Kong SAR, China\\
\medskip
}
\end{flushleft}

\clearpage

\setcounter{figure}{0}
\renewcommand\thefigure{S\arabic{figure}}
\setcounter{table}{0}
\renewcommand\thetable{S\arabic{table}}

\begin{center}
{\bf Supplemental Materials}
\end{center}

\noindent{\bf DAMPE experiment}\\
\noindent The DAMPE satellite was launched into a 500-km Sun-synchronous orbit on December 17, 2015, and has operated stably in space since then. The on-orbit calibration of each detector shows that the payload has maintained a good long-term performance over the past years\cite{DAMPE:2019lxv}. Also, the detector performance was validated with test beams of protons, electrons and fragmented ions with $A/Z=2$ from helium to argon. The linearity of the energy measurement of DAMPE is validated with the electron beams up to 243 GeV\cite{Zhang:2016xkz}. At higher energies, the laser test suggested that the measurement of the deposited energy in a single BGO crystal retains a good linearity up to a few TeV\cite{Zhao:2022wfw}. The absolute energy scale of DAMPE is determined with the geomagnetic cutoff of the electron plus positron spectrum, which results in a small bias of $\sim$1.3\%\cite{Zang:2025erm} and is not corrected.

\noindent{\bf Monte Carlo simulation}\\
\noindent Extensive MC simulations are carried out to obtain the instrument response to incident particles in the DAMPE detector. The GEometry ANd Tracking (GEANT) toolkit v4.10.05\cite{GEANT4:2002zbu} with the FTFP\_BERT physics list is adopted for the simulations of nuclei up to 100 TeV/n. For higher energies, we use the EPOS\_LHC model from the the Cosmic Ray Monte Carlo (CRMC; https://web.ikp.kit.edu/rulrich/crmc.html) linked to GEANT through a custom interface\cite{Tykhonov:2021xlb}. The energy response of MC simulations is corrected by including the Birks' quenching effect when calculating the ionization energy deposits in the calorimeter\cite{Chen:2023koi}. Such a correction is important for high $Z$ and low energy particles. 

The simulated events are generated assuming an isotropic source with an $E^{-1}$ spectrum. The simulation data are re-weighted to $E^{-2.6}$ and $E^{-3.0}$ spectra, for primary (e.g. carbon, oxygen, and iron) and secondary (e.g. boron) nuclei, respectively. To evaluate the uncertainties from the hadronic interaction model, another simulation program is developed with the FLUKA\cite{Bohlen:2014buj} 2011.2x package, which uses DPMJET3 for nucleus-nucleus interaction above 5 GeV/n. 

\noindent{\bf Event selections}\\
\noindent Given that the analysis procedures for protons and helium nuclei closely follow those established in refs.\cite{DAMPE:2019gys} and \cite{Alemanno:2021gpb}, the present work focuses primarily on the analysis of carbon, oxygen, and iron nuclei. Nine years of on-orbit data are used with a total exposure time of $\sim 2.17 \times 10^{8}$\,s, corresponding to $\sim$76.2\% of the operation time. A general pre-selection is applied to select events with good quality of reconstruction. The selection logic for different nuclei is broadly consistent, although the specific parameters are adjusted accordingly.

\begin{enumerate}
  \item[(I)] \textbf{Pre-selection.} The pre-selection is to ensure good shower development and proper event reconstruction in the calorimeter, including: 
  \begin{enumerate}
    \item[(i)] the total deposited energy ($E_{\rm dep}$) must exceed 50 GeV, 63 GeV, and 150 GeV for carbon, oxygen, and iron nuclei, respectively, to minimize the geomagnetic cutoff effect\cite{Smart:2005abc};
  
    \item[(ii)] the energy deposited in any single BGO layer must not exceed 35\% of the total deposited energy to exclude events incident from the sides.
   
  \end{enumerate}
  
    \item[(II)] \textbf{Trigger condition.} Events satisfying the High Energy Trigger (HET) condition\cite{DAMPE:2017cev} are selected. The HET is optimized to select well-contained events. For carbon, oxygen, and iron nuclei, the HET efficiency exceeds 95\%. 

  \item[(III)] \textbf{Track selection.} A deep learning based tracking approach\cite{Tykhonov:2022acr} is employed, which achieves a better performance especially at high energies than the Kalman filter track reconstruction. The reconstructed track is further required to align with the maximum number of hits in both the PSD-$x$ and PSD-$y$ layers and to penetrate the calorimeter from top to bottom. This ensures that the event is fully contained within the detector and the key characteristics can be accurately measured.

  \item[(IV)] \textbf{Charge selection.} The particle charge $Z$ is measured by both the PSD and the STK. For carbon and oxygen, the first STK layer charge is required to satisfy $Q_{\mathrm{STK1}} > 2.8$ to suppress contamination from protons and helium nuclei. For PSD charge measurement, the signals from the four sub-layers are employed to build a global PSD charge, with proper corrections due to strip alignments, light attenuation, fluorescence quenching, and equalization of different channels\cite{Dong:2018qof,Ma:2018brb}. The candidate events are selected using energy-dependent charge intervals as:  
\begin{align*}
  5.7 &< Q_{\mathrm{PSD}} < 6.3 + 0.015\cdot \log^2(E_{\rm dep}/{\rm GeV}), \quad (\text{for carbon}), \\
  7.7 &< Q_{\mathrm{PSD}} < 8.35 + 0.015\cdot \log^2(E_{\rm dep}/{\rm GeV}), \quad (\text{for oxygen}), \\
  25.5 &< Q_{\mathrm{PSD}} < 27.2  \quad (\text{for iron}).
\end{align*}
The charge selection criteria maintain an efficiency of approximately 90\% over the full analysis energy range. The same procedure and selections are applied to the MC simulations, and the MC charge distributions are shrinked to match the flight data for different particles. For proton and helium analysis, a combination of the PSD and STK charge measurements is employed to keep a stable selection efficiency at  high energies ($>10$ TeV).

\end{enumerate}

\begin{figure}[!ht]
\begin{center}
\includegraphics[width=0.48\columnwidth]{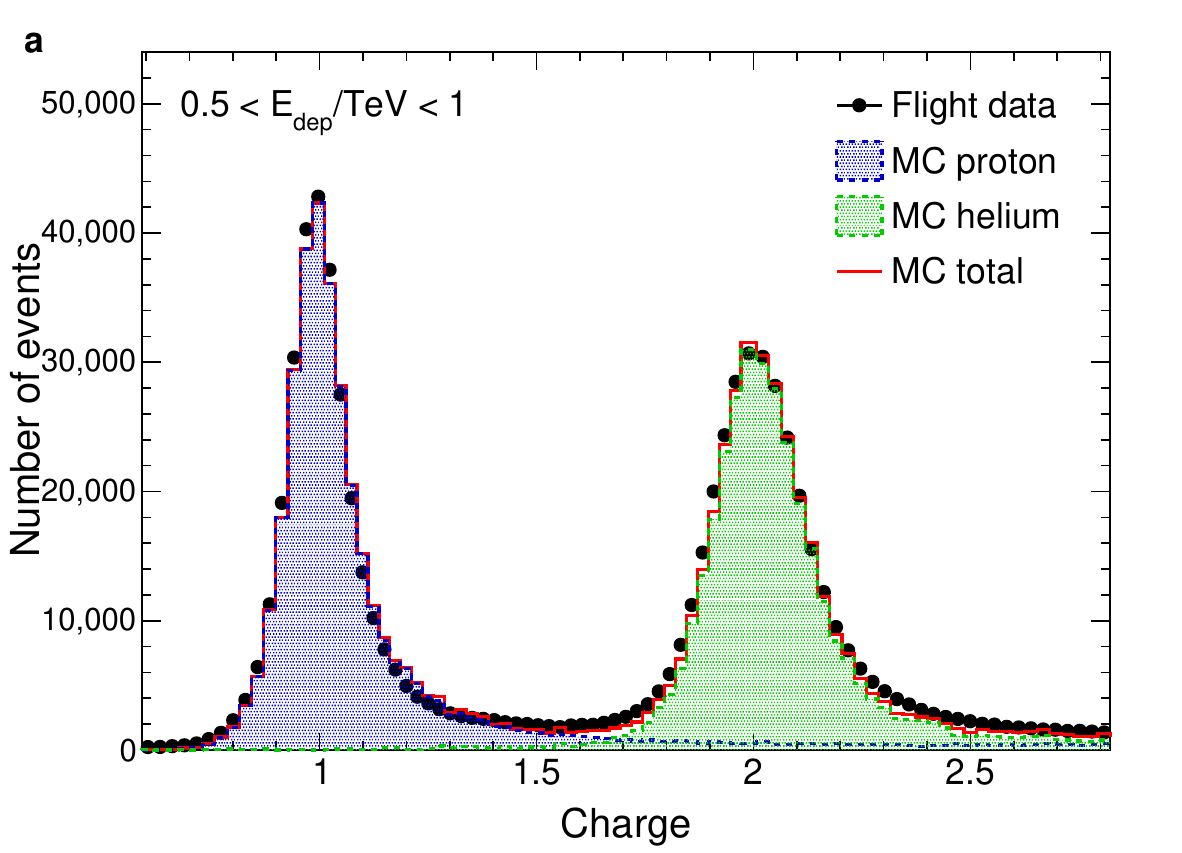}
\includegraphics[width=0.48\columnwidth]{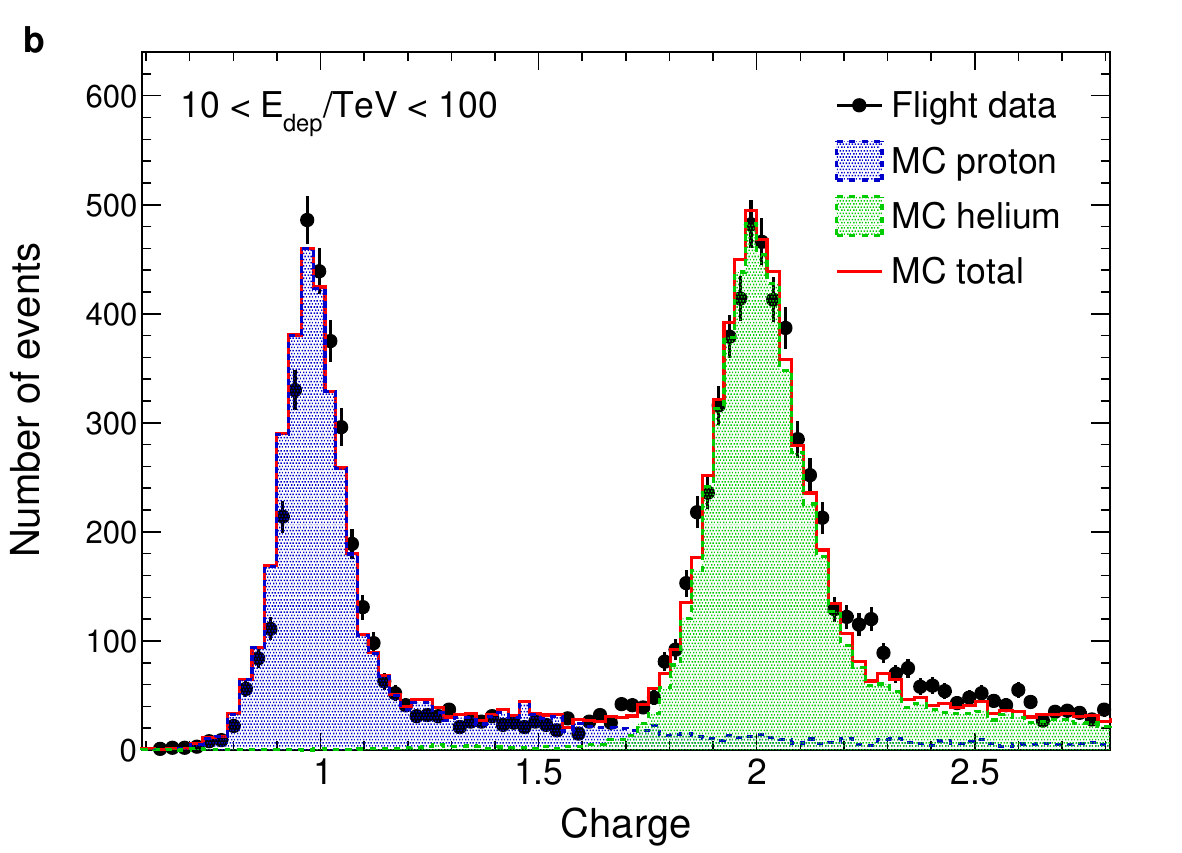}
\includegraphics[width=0.48\columnwidth]{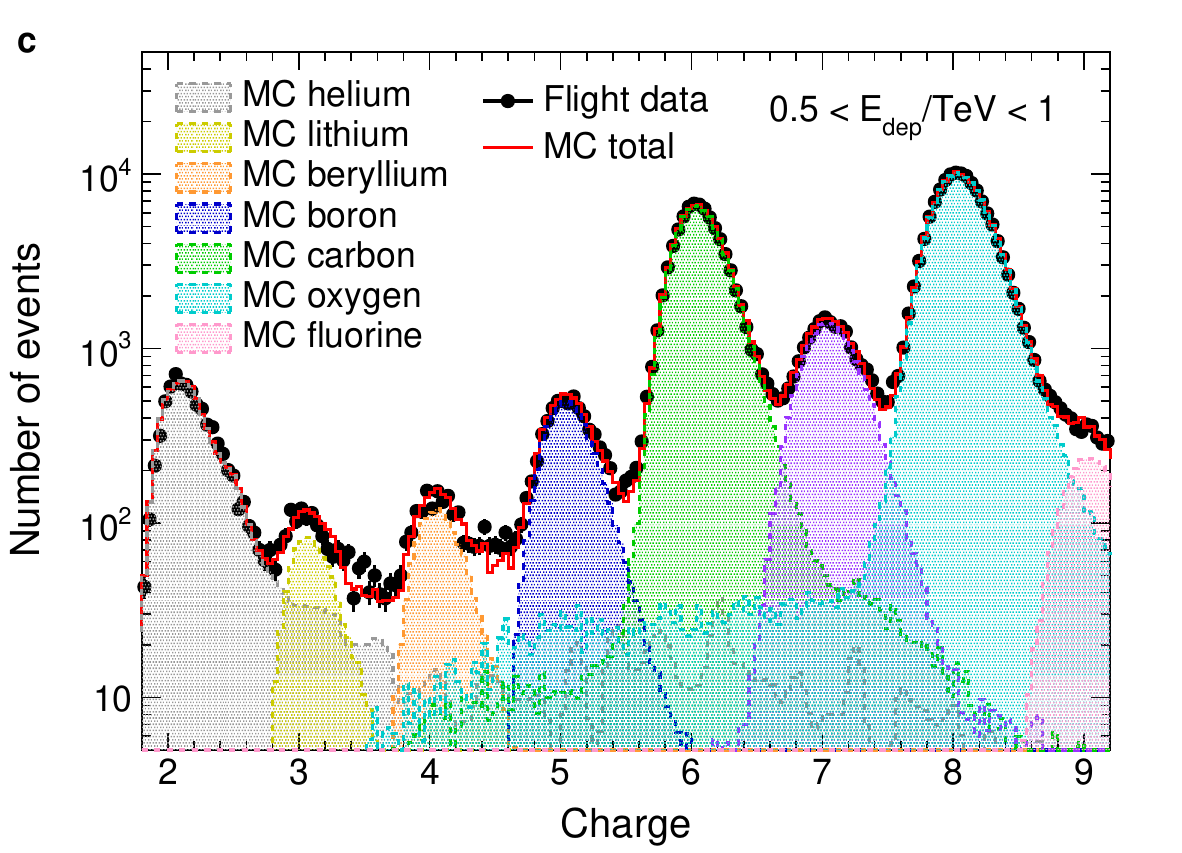}
\includegraphics[width=0.48\columnwidth]{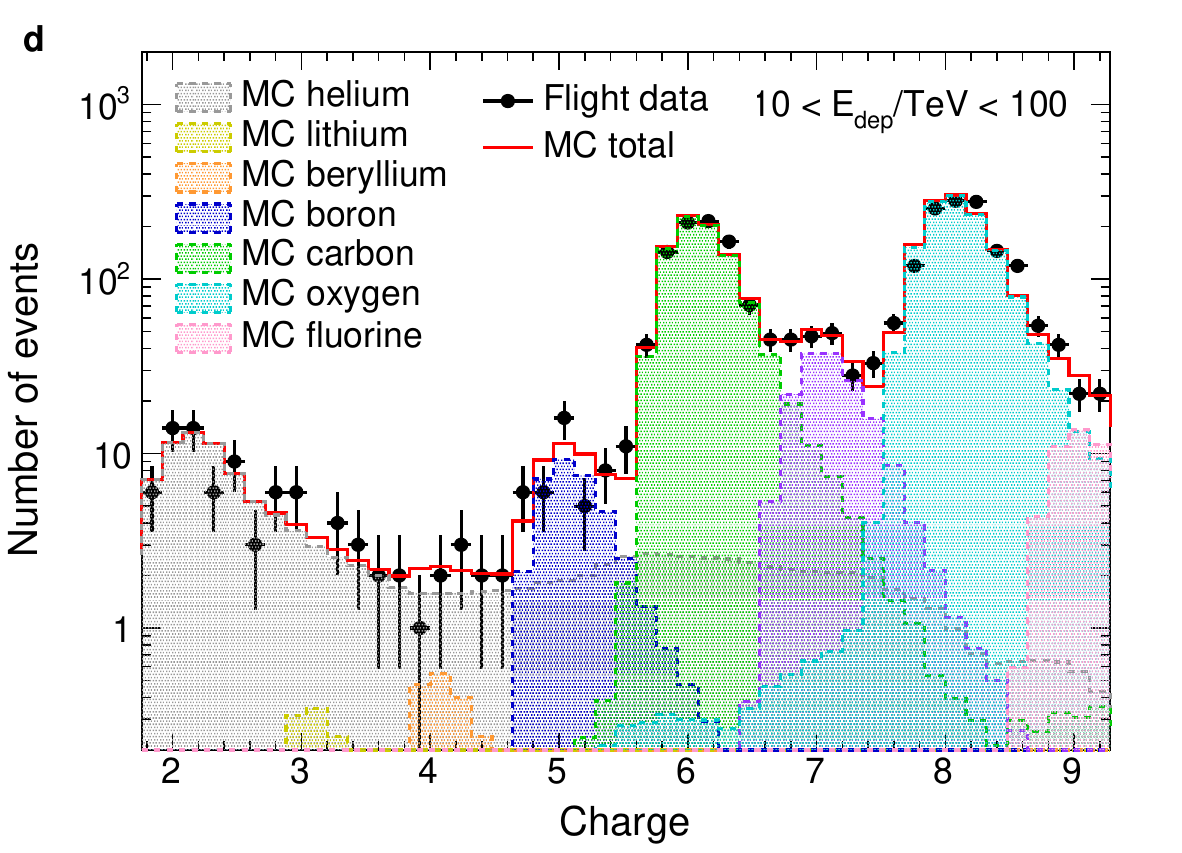}
\includegraphics[width=0.48\columnwidth]{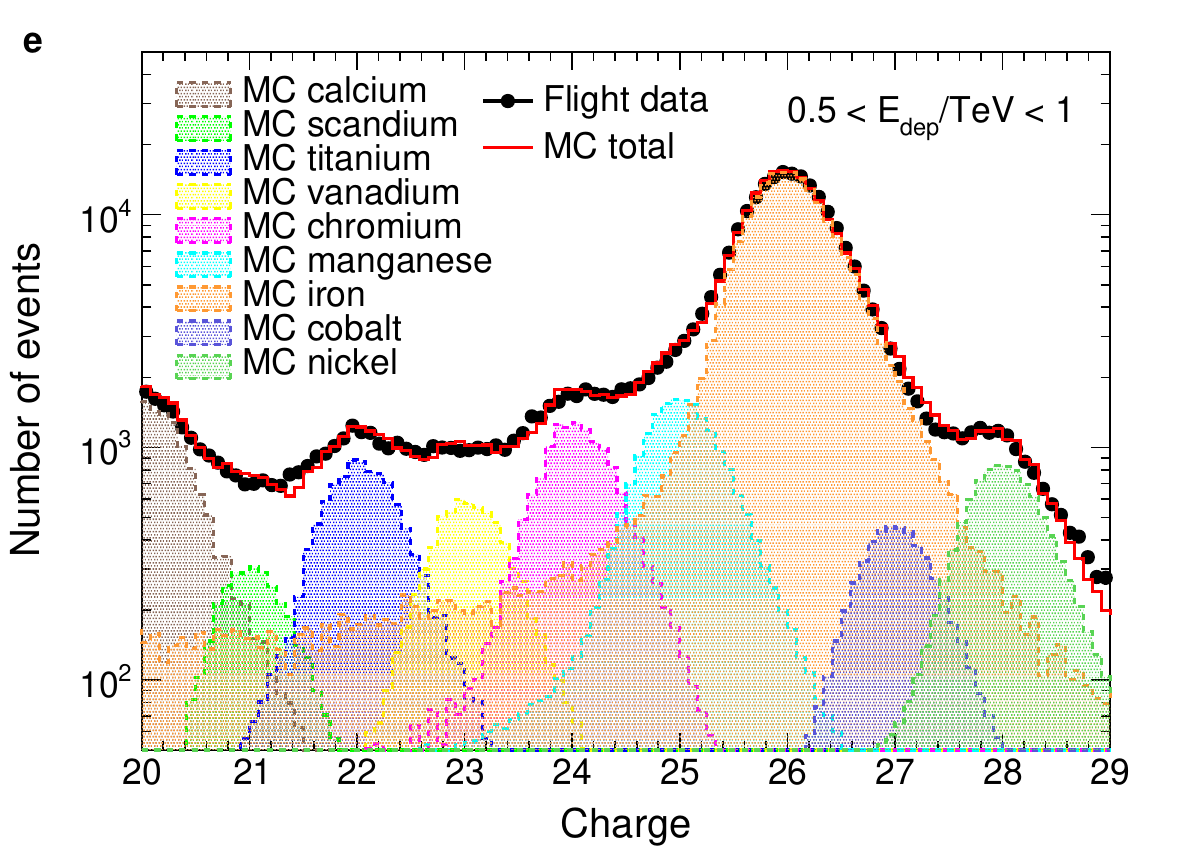}
\includegraphics[width=0.48\columnwidth]{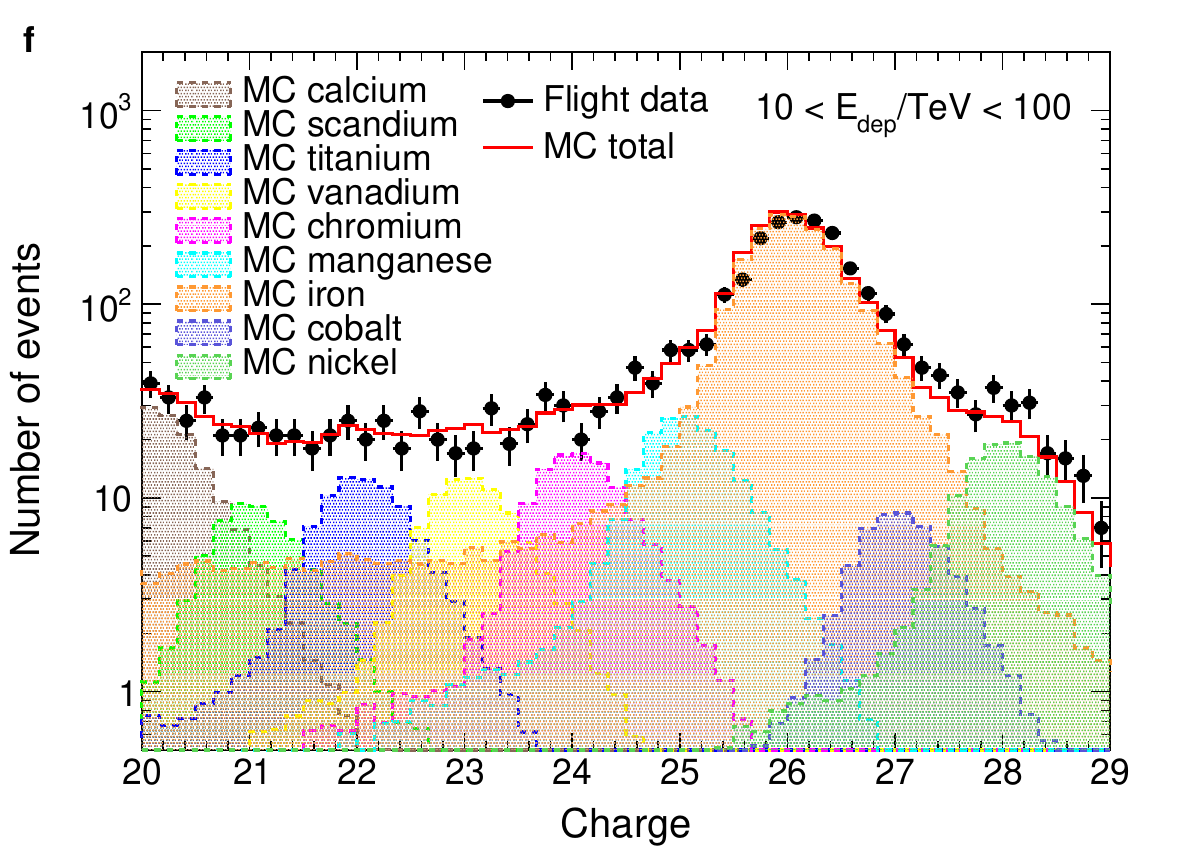}
\end{center}
\caption{{\bf Charge distributions of events.}
The left panels ({\bf a}, {\bf c}, {\bf e}) are for deposited energy bin of $0.5<E_{\rm dep}<1$ TeV, 
and the right panels ({\bf b}, {\bf d}, {\bf f}) are for deposited energy bin of $10<E_{\rm dep}<100$ TeV.
From top to bottom, three charge ranges relevant to protons and helium, carbon and oxygen, and iron nuclei
are shown. Black dots depict the flight data. Dashed lines with different colors show the best-fit 
MC simulated samples. The red solid lines show the sum of MC samples. 
}
\label{fig-temfit}
\end{figure}

\noindent\textbf{Background subtraction}\\
\noindent An MC-based template fit is performed on the charge distribution to estimate the background contamination within the selected charge windows for each species, as shown in as shown in Figure \ref{fig-temfit}. In these figures, the hatched regions show the \textsc{Geant4} simulated charge distributions of given particle species, and the red solid lines show the total contribution of relevant species. Good match between the data and the simulation can be obtained. 

The contamination for target nuclei are estimated through computing the fractions of adjacent nuclei in the specified charge window according to the MC templates.
For proton and helium, their cross-contamination dominate the backgrounds, which are estimated to be less than 4\% and 5\%, respectively, in the whole energy range. For carbon, in the lower-energy region, the dominant background arises from the fragmentation of heavier nuclei (particularly oxygen). 
At high energies, above the TeV scale, high-abundance helium nuclei can produce showers in the calorimeter that spawn back-scattered secondaries penetrating the PSD strips. This leads to an overestimation of the reconstructed charge and introduces a non-negligible contamination in both carbon and oxygen samples.
Consequently, the total contamination ratios for carbon and oxygen increase from 0.5\% and 0.4\% at 100 GeV to 2.4\% and 1.8\% at 50 TeV, respectively. For iron, both lighter nuclei (chromium, manganese) and heavier ones (e.g., cobalt and nickel) are far less abundant than iron itself. Moreover, the iron signal is sufficiently large that the impact of back-scattered shower particles on charge reconstruction is negligible. Therefore, the contamination ratio of the iron sample remains within 3\% in the whole energy range.

\noindent{\bf CR flux measurements}\\
\noindent The total deposited energy in the calorimeter ($E_{dep}$) is measured by summing up the energies of all BGO bars together. When the energy deposited in a single BGO bar exceeds $\sim 4$ TeV, some readout channels may saturate\cite{Yue:2020hmj}, leading to non-linearity in the energy response. A deep learning approach\cite{Serpolla:2025vrj} based on MC simulation was developed to correct the energy deposit(s) of those saturated bar(s) for nuclei up to PeV. Due to the relatively large spread of the calorimeter's energy response to the nuclei, an unfolding procedure\cite{DAgostini:1994fjx} is employed to derive the primary fluxes of CRs. The differential flux in a given incident energy bin $[E_i,E_i+\Delta E_i]$ is then calculated as
\begin{equation}
\Phi(E_{i},E_{i}+\Delta E_{i})= \frac{N_{{\rm inc},i}}
{\Delta E_i~A_{{\rm eff},i}~\Delta t},
\end{equation}
where $\Delta E_i$ is the energy bin width, $A_{{\rm eff},i}$ is the effective acceptance, 
and $\Delta t$ is the live time. The unfolded number of events, $N_{{\rm inc},i}$, is given by 
\begin{equation}
N_{{\rm inc},i}=\sum_{j=1}^n P(E_i|W_j)N_{{\rm dep},j},
\end{equation}
where $N_{{\rm dep},j}$ is the number of events in the $j$-th deposit energy bin
$[W_j,W_j+\Delta W_j]$, and $P(E_i|W_j)$ is the transfer probability from the $j$-th deposit
energy bin to the $i$-th incident energy bin.

The effective acceptance in the $i$-th kinetic energy bin is estimated from MC samples as
\begin{equation}
    A_{{\rm eff},i} = G_{\mathrm{gen}} \times \frac{N_i^{\mathrm{sel}}}{N_i^{\mathrm{gen}}},
\end{equation}
where $G_{\mathrm{gen}}$ is the geometric factor corresponding to the generation surface in the MC simulation, $N_i^{\mathrm{gen}}$ is the number of generated events in the $i$-th energy bin, and $N_i^{\mathrm{sel}}$ is the number of events that pass all of the selection criteria. 

\noindent{\bf Uncertainties}\\
\noindent The statistical uncertainties are associated with Poisson fluctuations of the number of 
detected events $N(E_{\rm dep,j})$. To properly determine the error propagation in the unfolding 
procedure, a toy-MC approach is applied, via sampling the event numbers in each deposited energy 
bin with Poisson fluctuations. The variations of the unfolded numbers of events in each kinetic energy 
bin are then obtained and the standard deviations are assigned as the 1$\sigma$ statistical uncertainty. 

The systematic uncertainties due to different sources are investigated extensively. The selection
efficiencies derived from the MC simulation, e.g. HET and track, are compared with the efficiencies
estimated from the flight data for different particles and the deviations are taken as the associated
systematic uncertainties. The uncertainty of charge selection is estimated by varying the charge 
selection window and checking the flux variations. The uncertainty due to background subtraction is
estimated by considering the errors of the contamination fractions from the template fit procedure. 
The uncertainty from the BGO saturation correction is obtained by artificially modifying the energy
correction for the saturated events by $\pm$1 standard deviation of the corrected ratio variation. 

For proton and helium, the related uncertainties are found to be significantly below 1\% up to 200 TeV 
and thus negligible in the current flux measurements. Finally, the systematic uncertainty associated 
with the hadronic interaction model used in the MC simulation is obtained through comparing simulation
results with the test beam data or two different simulation tools, i.e. GEANT4 and FLUKA. 
Specifically, for proton, helium, carbon and oxygen, the uncertainty of the hadronic interaction model 
in lower energy region ($<$ 400 GeV for proton and $<$ 75 GeV/n for others) is estimated according to 
the comparison of the energy response between the test beam data and the GEANT4 simulation. 
At higher energies, the uncertainty is assigned as the difference between the fluxes measured using 
the GEANT4 and the FLUKA simulations. For iron, no beam test data is available and hence the uncertainty 
is evaluated as the difference between the two simulations.

\begin{figure}[!ht]
\begin{center}
\includegraphics[width=0.48\columnwidth]{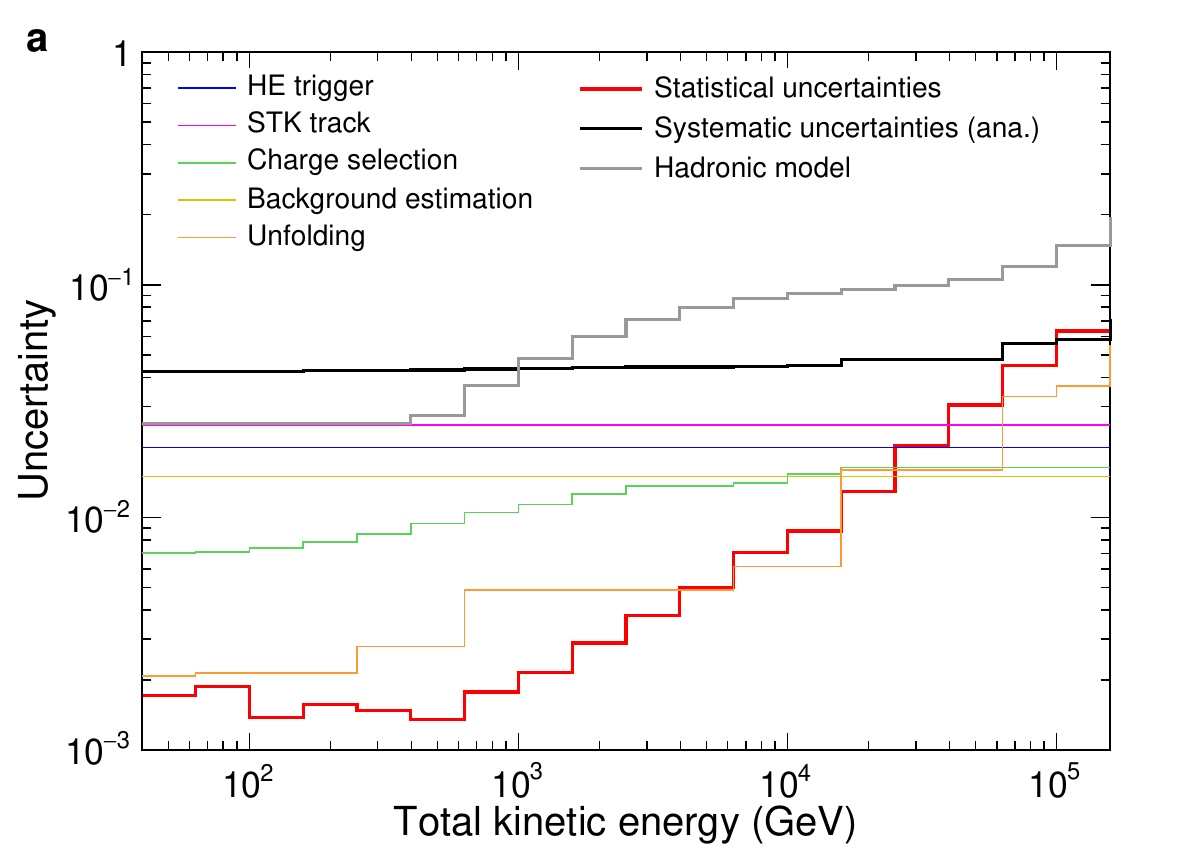}
\includegraphics[width=0.48\columnwidth]{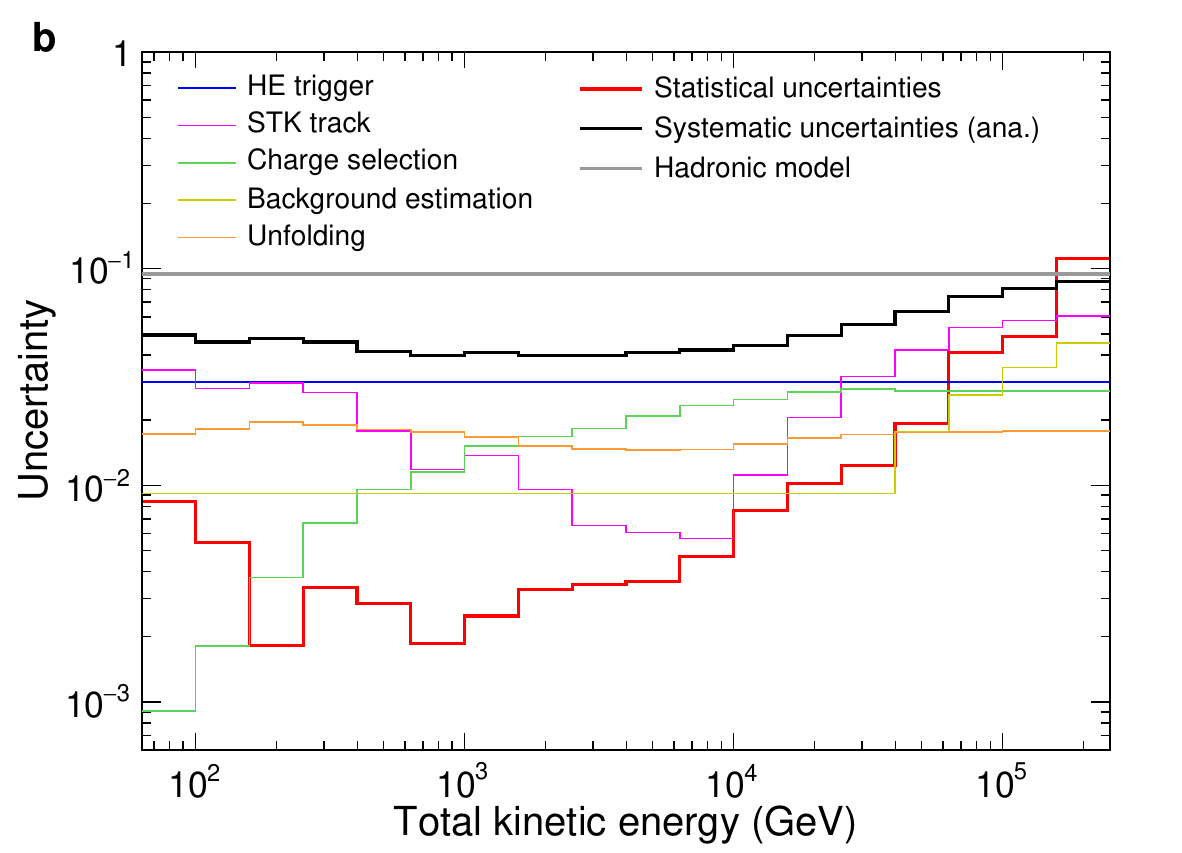}
\includegraphics[width=0.48\columnwidth]{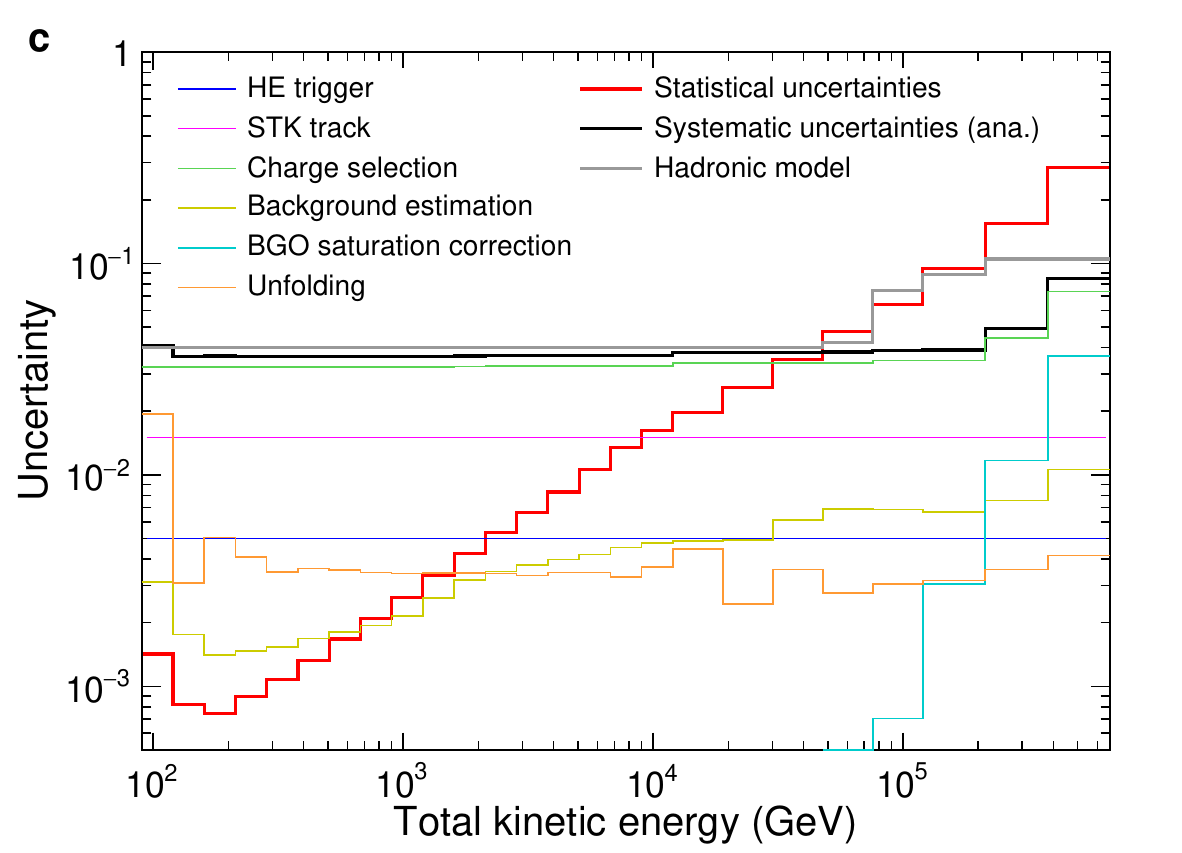}
\includegraphics[width=0.48\columnwidth]{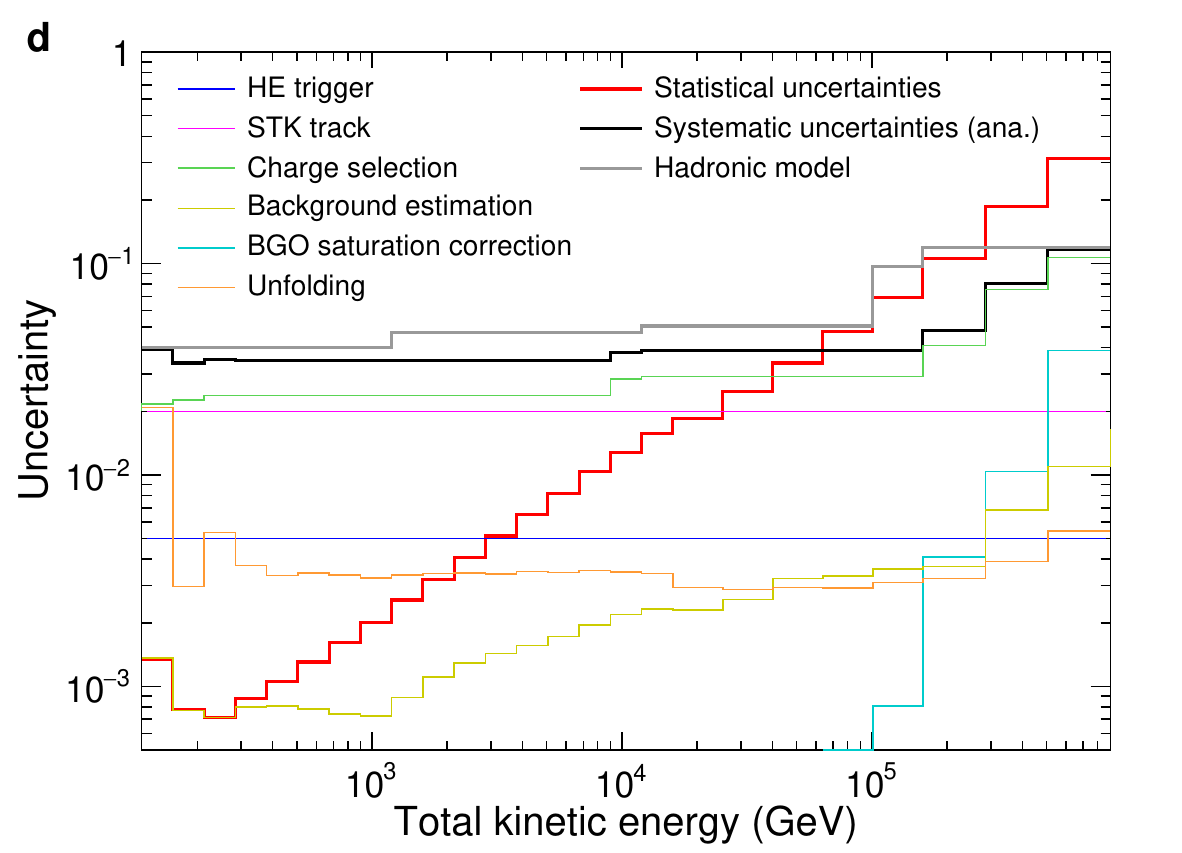}
\includegraphics[width=0.48\columnwidth]{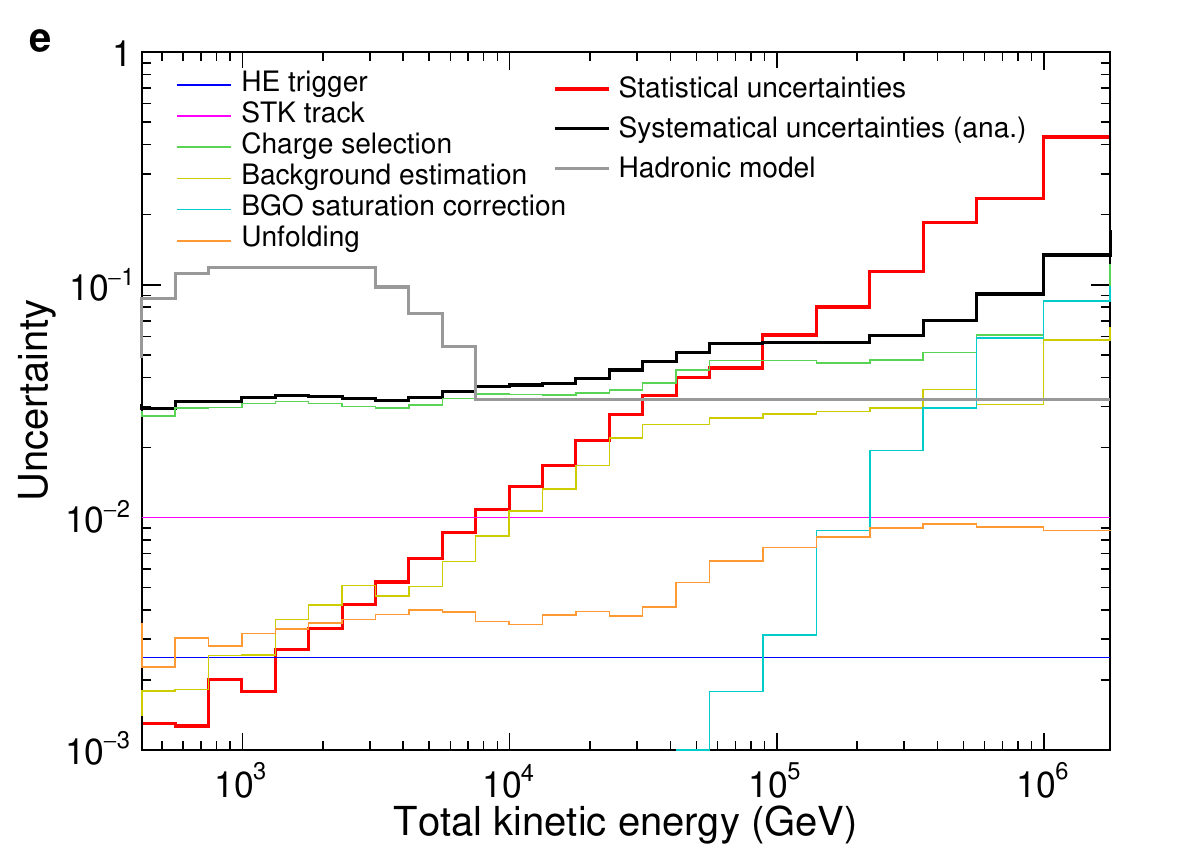}
\end{center}
\caption{{\bf Relative uncertainties versus energy.}
Panels {\bf a} - {\bf e} correspond to those of proton, helium, carbon, oxygen, and iron,
respectively. The systematic uncertainties in analysis, shown in black, are the quadratic sum of 
all systematic contributions except for those from the hadronic models.
} 
\label{fig-unc}
\end{figure}

The total uncertainty budgets for proton, helium, carbon, oxygen and iron are shown in Figure \ref{fig-unc}. The measured fluxes and the associated uncertainties, including the statistical uncertainties and systematic ones from the analysis procedure and the hadronic interaction models, are presented in Tables \ref{Table:p_flux_ek}, \ref{Table:He_flux_ek}, \ref{Table:C_flux_ek}, \ref{Table:O_flux_ek}, and \ref{Table:Fe_flux_ek}.

\begin{table*}[htbp]
\centering
\caption{The proton fluxes as function of total kinetic energy ($E_t$) measured by DAMPE, together with the 1$\sigma$ statistical errors and the systematic uncertainties from the analysis and hadronic interaction models.}
\begin{tabular}{ c | c }
\hline
  $E_t$ Bin & $\Phi \pm \sigma_{\rm stat}  \pm \sigma_{\rm sys}^{\rm ana} \pm \sigma_{\rm sys}^{\rm had}$ \\
  (GeV)& (GeV$^{-1}$~m$^{-2}$~s$^{-1}$~sr$^{-1}$) \\
\hline
39.8 - 63.1 & $( 2.81 \pm 0.005 \pm 0.12 \pm 0.07  ) \times 10^{-1} $ \\
63.1 - 100.0 & $( 8.05 \pm 0.015 \pm 0.34 \pm 0.20 ) \times 10^{-2} $ \\
100.0 - 158.5 & $( 2.27 \pm 0.003 \pm 0.10 \pm 0.06 ) \times 10^{-2} $ \\ 
158.5 - 251.2 & $( 6.36 \pm 0.010 \pm 0.27 \pm 0.16 ) \times 10^{-3} $ \\
251.2 - 398.1 & $( 1.81 \pm 0.003 \pm 0.08 \pm 0.05 ) \times 10^{-3} $ \\
398.1 - 631.0 & $( 5.23 \pm 0.007 \pm 0.22 \pm 0.14 ) \times 10^{-4} $ \\
631.0 - 1000 & $( 1.54 \pm 0.003 \pm 0.07 \pm 0.06 ) \times 10^{-4} $ \\
1000 - 1585 & $( 4.60 \pm 0.010 \pm 0.20 \pm 0.22 ) \times 10^{-5} $ \\
1585 - 2512& $( 1.39 \pm 0.004 \pm 0.06 \pm 0.08 ) \times 10^{-5} $ \\
2512 - 3981 & $( 4.24 \pm 0.015 \pm 0.19 \pm 0.30 ) \times 10^{-6} $ \\
3981 - 6310 & $( 1.30 \pm 0.006 \pm 0.06 \pm 0.11 ) \times 10^{-6} $ \\
6310 - 10000 & $( 3.99 \pm 0.027 \pm 0.18 \pm 0.35 ) \times 10^{-7} $ \\
10000 - 15849 & $( 1.18 \pm 0.010 \pm 0.05 \pm 0.11 ) \times 10^{-7} $ \\
15849 - 25119 & $( 3.34 \pm 0.043 \pm 0.16 \pm 0.32 ) \times 10^{-8} $ \\
25119 - 39811 & $( 8.85 \pm 0.190 \pm 0.42 \pm 0.88 ) \times 10^{-9} $ \\
39811 - 63096 & $( 2.30 \pm 0.073 \pm 0.11 \pm 0.25 ) \times 10^{-9} $ \\
63096 - 100000 & $( 6.05 \pm 0.281 \pm 0.34 \pm 0.73 ) \times 10^{-10} $ \\
100000 - 158489 & $( 1.69 \pm 0.109 \pm 0.10 \pm 0.25 ) \times 10^{-10} $ \\
\hline
\end{tabular}
\label{Table:p_flux_ek}
\end{table*}

\begin{table*}[htbp]
\centering
\caption{The helium fluxes as function of total kinetic energy ($E_t$) measured by DAMPE, together with the 1$\sigma$ statistical errors and the systematic uncertainties from the analysis and hadronic interaction models.}
\begin{tabular}{ c | c }
\hline
  $E_t$ Bin & $\Phi \pm \sigma_{\rm stat}  \pm \sigma_{\rm sys}^{\rm ana} \pm \sigma_{\rm sys}^{\rm had}$ \\
  (GeV)& (GeV$^{-1}$~m$^{-2}$~s$^{-1}$~sr$^{-1}$) \\
\hline
63.1 - 100.0 & $( 5.31 \pm 0.045 \pm 0.28 \pm 0.50 ) \times 10^{-2} $ \\
100.0 - 158.5 & $( 1.60 \pm 0.009 \pm 0.08 \pm 0.15 ) \times 10^{-2} $ \\ 
158.5 - 251.2 & $( 4.56 \pm 0.083 \pm 0.24 \pm 0.43 ) \times 10^{-3} $ \\
251.2 - 398.1 & $( 1.30 \pm 0.004 \pm 0.07 \pm 0.12 ) \times 10^{-3} $ \\
398.1 - 631.0 & $( 3.81 \pm 0.011 \pm 0.18 \pm 0.36 ) \times 10^{-4} $ \\
631.0 - 1000 & $( 1.13 \pm 0.002 \pm 0.05 \pm 0.11 ) \times 10^{-4} $ \\
1000 - 1585 & $( 3.42 \pm 0.009 \pm 0.16 \pm 0.32 ) \times 10^{-5} $ \\
1585 - 2512 & $( 1.05 \pm 0.003 \pm 0.05 \pm 0.10 ) \times 10^{-5} $ \\
2512 - 3981 & $( 3.31 \pm 0.012 \pm 0.15 \pm 0.31 ) \times 10^{-6} $ \\
3981 - 6310 & $( 1.06 \pm 0.004 \pm 0.05 \pm 0.10 ) \times 10^{-6} $ \\
6310 - 10000 & $( 3.47 \pm 0.016 \pm 0.16 \pm 0.33 ) \times 10^{-7} $ \\
10000 - 15849 & $( 1.16 \pm 0.009 \pm 0.06 \pm 0.11 ) \times 10^{-7} $ \\
15849 - 25119 & $( 3.82 \pm 0.039 \pm 0.20 \pm 0.36 ) \times 10^{-8} $ \\
25119 - 39811 & $( 1.21 \pm 0.015 \pm 0.07 \pm 0.11 ) \times 10^{-8} $ \\
39811 - 63096 & $( 3.57 \pm 0.069 \pm 0.24 \pm 0.34 ) \times 10^{-9} $ \\
63096 - 100000 & $( 1.01 \pm 0.042 \pm 0.08 \pm 0.10 ) \times 10^{-9} $ \\
100000 - 158489 & $( 2.85 \pm 0.139 \pm 0.24 \pm 0.27 ) \times 10^{-10} $ \\
158489 - 251189 & $( 8.52 \pm 0.954 \pm 0.78 \pm 0.81 ) \times 10^{-11} $ \\
\hline
\end{tabular}
\label{Table:He_flux_ek}
\end{table*}

\begin{table*}[htbp]
\centering
\caption{The carbon fluxes as function of total kinetic energy ($E_t$) measured by DAMPE, together with the 1$\sigma$ statistical errors and the systematic uncertainties from the analysis and hadronic interaction models.}
\begin{tabular}{ c | c }
\hline
  $E_t$ Bin & $\Phi \pm \sigma_{\rm stat}  \pm \sigma_{\rm sys}^{\rm ana} \pm \sigma_{\rm sys}^{\rm had}$ \\
  (GeV)& (GeV$^{-1}$~m$^{-2}$~s$^{-1}$~sr$^{-1}$) \\
\hline
90.0 - 120.0 & $( 4.60 \pm 0.007 \pm 0.19 \pm 0.18  ) \times 10^{-3} $ \\
120.0 - 160.0 & $( 2.29 \pm 0.002 \pm 0.08 \pm 0.09  ) \times 10^{-3} $ \\
160.0 - 213.4 & $( 1.11 \pm 0.001 \pm 0.04 \pm 0.04  ) \times 10^{-3} $ \\
213.4 - 284.6 & $( 5.26 \pm 0.005 \pm 0.19 \pm 0.21  ) \times 10^{-4} $ \\
284.6 - 379.5 & $( 2.48 \pm 0.003 \pm 0.09 \pm 0.10  ) \times 10^{-4} $ \\
379.5 - 506.0 & $( 1.17 \pm 0.002 \pm 0.04 \pm 0.05  ) \times 10^{-4} $ \\
506.0 - 674.8 & $( 5.44 \pm 0.009 \pm 0.20 \pm 0.22  ) \times 10^{-5} $ \\
674.8 - 899.9 & $( 2.51 \pm 0.005 \pm 0.09 \pm 0.10  ) \times 10^{-5} $ \\
899.9 - 1200 & $( 1.17 \pm 0.003 \pm 0.04 \pm 0.05  ) \times 10^{-5} $ \\
1200 - 1600 & $( 5.43 \pm 0.018 \pm 0.20 \pm 0.22  ) \times 10^{-6} $ \\
1600 - 2134 & $( 2.53 \pm 0.011 \pm 0.09 \pm 0.10  ) \times 10^{-6} $ \\
2134 - 2846 & $( 1.18 \pm 0.006 \pm 0.04 \pm 0.05  ) \times 10^{-6} $ \\
2846 - 3795 & $( 5.54 \pm 0.037 \pm 0.20 \pm 0.22  ) \times 10^{-7} $ \\
3795 - 5060 & $( 2.64 \pm 0.022 \pm 0.10 \pm 0.11  ) \times 10^{-7} $ \\
5060 - 6748 & $( 1.27 \pm 0.014 \pm 0.05 \pm 0.05  ) \times 10^{-7} $ \\
6748 - 8999 & $( 6.07 \pm 0.082 \pm 0.22 \pm 0.24  ) \times 10^{-8} $ \\
8999 - 12000 & $( 2.96 \pm 0.049 \pm 0.11 \pm 0.12  ) \times 10^{-8} $ \\
12000 - 19019 & $( 1.20 \pm 0.024 \pm 0.05 \pm 0.05  ) \times 10^{-8} $ \\
19019 - 30143 & $( 3.86 \pm 0.101 \pm 0.15 \pm 0.16  ) \times 10^{-9} $ \\
30143 - 47773 & $( 1.29 \pm 0.046 \pm 0.05 \pm 0.05  ) \times 10^{-9} $ \\
47773 - 75715 & $( 4.62 \pm 0.022 \pm 0.18 \pm 0.20  ) \times 10^{-10} $ \\
75715 - 120000 & $( 1.52 \pm 0.098 \pm 0.06 \pm 0.11  ) \times 10^{-10} $ \\
120000 - 213394 & $( 3.69 \pm 0.350 \pm 0.14 \pm 0.33  ) \times 10^{-11} $ \\
213394 - 379474 & $( 7.37 \pm 1.135 \pm 0.36 \pm 0.78  ) \times 10^{-12} $ \\
379474 - 674809 & $( 1.48 \pm 0.421 \pm 0.12 \pm 0.16  ) \times 10^{-12} $ \\
\hline
\end{tabular}
\label{Table:C_flux_ek}
\end{table*}

\begin{table*}[htbp]
\centering
\caption{The oxygen fluxes as function of total kinetic energy ($E_t$) measured by DAMPE, together with the 1$\sigma$ statistical errors and the systematic uncertainties from the analysis and hadronic interaction models.}
\begin{tabular}{ c | c }
\hline
  $E_t$ Bin & $\Phi \pm \sigma_{\rm stat}  \pm \sigma_{\rm sys}^{\rm ana} \pm \sigma_{\rm sys}^{\rm had}$ \\
  (GeV)& (GeV$^{-1}$~m$^{-2}$~s$^{-1}$~sr$^{-1}$) \\
\hline
120.0 - 160.0 & $( 3.47 \pm 0.005 \pm 0.14 \pm 0.14  ) \times 10^{-3} $ \\
160.0 - 213.4 & $( 1.77 \pm 0.001 \pm 0.06 \pm 0.07  ) \times 10^{-3} $ \\
213.4 - 284.5 & $( 8.66 \pm 0.006 \pm 0.30 \pm 0.35  ) \times 10^{-4} $ \\
284.5 - 379.4 & $( 4.19 \pm 0.004 \pm 0.15 \pm 0.17  ) \times 10^{-4} $ \\
379.4 - 506.0 & $( 2.01 \pm 0.002 \pm 0.07 \pm 0.08  ) \times 10^{-4} $ \\
506.0 - 674.7 & $( 9.47 \pm 0.012 \pm 0.33 \pm 0.38  ) \times 10^{-5} $ \\
674.7 - 899.7 & $( 4.46 \pm 0.007 \pm 0.15 \pm 0.18  ) \times 10^{-5} $ \\
899.7 - 1200 & $( 2.08 \pm 0.004 \pm 0.07 \pm 0.08  ) \times 10^{-5} $ \\
1200 - 1600 & $( 9.66 \pm 0.025 \pm 0.33 \pm 0.46  ) \times 10^{-6} $ \\
1600 - 2134 & $( 4.50 \pm 0.014 \pm 0.16 \pm 0.21  ) \times 10^{-6} $ \\
2134 - 2845 & $( 2.09 \pm 0.008 \pm 0.07 \pm 0.10  ) \times 10^{-6} $ \\
2845 - 3794 & $( 9.69 \pm 0.050 \pm 0.34 \pm 0.46  ) \times 10^{-7} $ \\
3794 - 5060 & $( 4.51 \pm 0.029 \pm 0.16 \pm 0.21  ) \times 10^{-7} $ \\
5060 - 6747 & $( 2.13 \pm 0.017 \pm 0.07 \pm 0.10  ) \times 10^{-7} $ \\
6747 - 8997 & $( 1.03 \pm 0.010 \pm 0.04 \pm 0.05  ) \times 10^{-7} $ \\
8997 - 11998 & $( 5.00 \pm 0.064 \pm 0.19 \pm 0.24  ) \times 10^{-8} $ \\
11998 - 16000 & $( 2.46 \pm 0.038 \pm 0.09 \pm 0.12  ) \times 10^{-8} $ \\
16000 - 25358 & $( 9.88 \pm 0.182 \pm 0.38 \pm 0.50  ) \times 10^{-9} $ \\
25358 - 40190 & $( 3.21 \pm 0.080 \pm 0.12 \pm 0.16  ) \times 10^{-9} $ \\
40190 - 63697 & $( 1.04 \pm 0.035 \pm 0.04 \pm 0.05  ) \times 10^{-9} $ \\
63697 - 100953 & $( 3.37 \pm 0.162 \pm 0.13 \pm 0.17  ) \times 10^{-10} $ \\
100953 - 160000 & $( 1.04 \pm 0.071 \pm 0.04 \pm 0.10  ) \times 10^{-10} $ \\
160000 - 284525 & $( 2.39 \pm 0.252 \pm 0.12 \pm 0.28  ) \times 10^{-11} $ \\
284525 - 505965 & $( 4.34 \pm 0.810 \pm 0.35 \pm 0.52  ) \times 10^{-12} $ \\
505965 - 899746 & $( 7.88 \pm 0.248 \pm 0.92 \pm 0.94  ) \times 10^{-13} $ \\
\hline
\end{tabular}
\label{Table:O_flux_ek}
\end{table*}

\begin{table*}[htbp]
\centering
\caption{The iron fluxes as function of total kinetic energy ($E_t$) measured by DAMPE, together with the 1$\sigma$ statistical errors and the systematic uncertainties from the analysis and hadronic interaction models.}
\begin{tabular}{ c | c }
\hline
  $E_t$ Bin & $\Phi \pm \sigma_{\rm stat}  \pm \sigma_{\rm sys}^{\rm ana} \pm \sigma_{\rm sys}^{\rm had}$ \\
  (GeV) & (GeV$^{-1}$~m$^{-2}$~s$^{-1}$~sr$^{-1}$) \\
\hline
419.9 - 560.0 & $( 1.21 \pm 0.002 \pm 0.04 \pm 0.11  ) \times 10^{-4} $ \\
560.0 - 746.8 & $( 6.32 \pm 0.008 \pm 0.20 \pm 0.71  ) \times 10^{-5} $ \\
746.8 - 995.8 & $( 3.21 \pm 0.006 \pm 0.10 \pm 0.38  ) \times 10^{-5} $ \\
995.8 - 1328 & $( 1.62 \pm 0.003 \pm 0.05 \pm 0.19  ) \times 10^{-5} $ \\
1328 - 1771 & $( 7.99 \pm 0.022 \pm 0.27 \pm 0.95  ) \times 10^{-6} $ \\
1771 - 2362 & $( 3.89 \pm 0.013 \pm 0.13 \pm 0.46  ) \times 10^{-6} $ \\
2362 - 3149 & $( 1.89 \pm 0.008 \pm 0.06 \pm 0.22  ) \times 10^{-6} $ \\
3149 - 4199 & $( 9.01 \pm 0.047 \pm 0.30 \pm 0.88  ) \times 10^{-7} $ \\
4199 - 5600 & $( 4.20 \pm 0.028 \pm 0.13 \pm 0.32  ) \times 10^{-7} $ \\
5600 - 7468 & $( 1.95 \pm 0.017 \pm 0.07 \pm 0.11  ) \times 10^{-7} $ \\
7468 - 9958 & $( 9.06 \pm 0.098 \pm 0.33 \pm 0.29  ) \times 10^{-8} $ \\
9958 - 13280 & $( 4.37 \pm 0.059 \pm 0.16 \pm 0.14  ) \times 10^{-8} $ \\
13280 - 17709 & $( 2.12 \pm 0.035 \pm 0.08 \pm 0.07  ) \times 10^{-8} $ \\
17709 - 23615 & $( 1.01 \pm 0.022 \pm 0.04 \pm 0.03  ) \times 10^{-8} $ \\
23615 - 31491 & $( 4.75 \pm 0.132 \pm 0.21 \pm 0.15  ) \times 10^{-9} $ \\
31491 - 41994 & $( 2.33 \pm 0.078 \pm 0.10 \pm 0.07  ) \times 10^{-9} $ \\
41994 - 56000 & $( 1.17 \pm 0.047 \pm 0.06 \pm 0.04  ) \times 10^{-9} $ \\
56000 - 88754 & $( 4.69 \pm 0.206 \pm 0.27 \pm 0.15  ) \times 10^{-10} $ \\
88754 - 140666 & $( 1.47 \pm 0.089 \pm 0.09 \pm 0.05  ) \times 10^{-10} $ \\
140666 - 222940 & $( 4.84 \pm 0.388 \pm 0.27 \pm 0.15  ) \times 10^{-11} $ \\
222940 - 353336 & $( 1.56 \pm 0.177 \pm 0.10 \pm 0.05  ) \times 10^{-11} $ \\
353336 - 560000 & $( 4.30 \pm 0.795 \pm 0.38 \pm 0.14  ) \times 10^{-12} $ \\
560000 - 995837 & $( 8.84 \pm 2.068 \pm 0.95 \pm 0.28  ) \times 10^{-13} $ \\
995837 - 1770876 & $( 1.32 \pm 0.589 \pm 0.19 \pm 0.04  ) \times 10^{-13} $ \\
\hline
\end{tabular}
\label{Table:Fe_flux_ek}
\end{table*}

\noindent{\bf Spectral fitting}\\
\noindent The significance of the spectral hardening or softening feature is estimated by comparing the 
fittings with two different models. One is a power-law (PL) function
\begin{equation}
    \Phi_{\rm PL}(E) = \Phi_0 \cdot \left( \frac{E}{10^3~\text{GeV}} \right)^{-\gamma},
\label{eq:pl}
\end{equation}
and the other is a Smoothly Broken Power Law (SBPL) function
\begin{equation}
    \Phi_{\rm SBPL}(E) = \Phi_0 \cdot \left( \frac{E}{10^3~\text{GeV}} \right)^{-\gamma} \left[ 1 + \left( \frac{E}{E_{\rm br}} \right)^s \right]^{\frac{\Delta \gamma}{s}},
\label{eq:sbpl}
\end{equation}
where $\gamma$ is the spectral index below the break energy $E_{\rm br}$, $\Delta\gamma$ is the 
change of the spectral index and and $s$ is the smoothness parameter.

The same fit method as refs.\cite{DAMPE:2019gys} and \cite{Alemanno:2021gpb} is applied to account for the systematic uncertainties by multiplying a set of independent nuisance parameters $w_j$ on the input model\cite{Fermi-LAT:2017bpc}. The $\chi^2$ function is
defined as
\begin{equation}
\chi^2=\sum_{i}\left[\frac{\Phi(E_{i})S(E_i;\,\boldsymbol{w})-\Phi_i}
{\sigma_{\rm stat,i}}\right]^2 + \sum_{j=1}^{m} \left(\frac{1-w_j}
{\tilde{\sigma}_{\rm sys,j}}\right)^2,
\end{equation}
where $\Phi_i$ and $\sigma_{\rm stat,i}$ are the flux and statistical uncertainty 
of the measurement in the $i$th kinetic energy bin, $\Phi(E_i)$ is the predicted flux 
in corresponding energy bin, $S(E_i;\,\boldsymbol{w})$ is a piecewise function defined 
by its value $w_j$ in corresponding energy range covered by the $j$th nuisance parameter, 
and $\tilde{\sigma}_{\rm sys,j}=\sqrt{\sigma_{\rm ana}^2+\sigma_{\rm had}^2}/\Phi$
is the relative systematic uncertainty of the data in such an energy range.
When calculating $\Phi(E_i)$, the flux at the central energy of each bin is used,
which results in negligible differences of the fitting parameters compared with the
method of integrating the fluxes over the energy bin.

The fit parameters of the SBPL model for the hardening and softening features are presented in Tables \ref{Table:fit_para_hard} and \ref{Table:fit_para_soft}, respectively. For the fit for the softening features, the smoothness parameter $s$ is fixed as $5$ or $10$ due to a lack of good constraint on it. The significances of those spectral features are evaluated by the comparisons of the $\chi^2$ values for the PL and SBPL models. For the hardening, the reduction of the $\chi^2/{\rm dof}$ is $867.4/3$, $562.2/3$, $128.5/3$, $110.1/3$ and $11.4/2$, corresponding to significance of $29\sigma$, $23\sigma$, $11\sigma$, $10\sigma$, and $2.7\sigma$ for proton, helium, carbon, oxygen, and iron, respectively. While, for the softening, the reduction of the $\chi^2/{\rm dof}$ is $102.5/2$, $58.2/2$, $14.7/2$, $21.6/2$ and $9.6/2$, corresponding to significance of $9.1\sigma$, $7.2\sigma$, $3.2\sigma$, $4.1\sigma$, and $2.4\sigma$ for proton, helium, carbon, oxygen, and iron, respectively.

\begin{table*}[htbp]
\centering
\caption{The parameters of the SBPL model obtained from the fit of the spectral hardening features for proton, helium, carbon, oxygen and iron.}
\resizebox{\columnwidth}{!}{
\begin{tabular}{c|ccccc}
\hline
\hline
Particle & Proton & Helium & Carbon & Oxygen & Iron \\
\hline
Fit energy range  & 0.063 - 6.3 TeV & 0.100 - 15.8 TeV & 0.506 - 47.8 TeV &  0.675 - 63.7 TeV & 1.77 - 223 TeV  \\ 
\hline
Nuisance parameters & 4 & 4 & 4 & 4 & 4 \\ 
\hline
$\Phi_0$ ($10^{-5}$ GeV$^{-1}$ m$^{-2}$ s$^{-1}$ sr$^{-1}$) & 7.11 $\pm$ 0.26
& 5.57 $\pm$ 0.34 & 1.28 $\pm$ 0.04 & 2.27 $\pm$ 0.07 & 2.55 $\pm$ 0.11 \\
$E_{\rm br}$ (TeV) & 0.59 $\pm$ 0.04 & 1.19 $\pm$ 0.08 & 5.35 $\pm$ 1.26  
& 6.39 $\pm$ 0.61 & 28.7 $\pm$ 11.3 \\ 
$\gamma$ & 2.77 $\pm$ 0.01 & 2.72 $\pm$ 0.01 & 2.68 $\pm$ 0.01 
& 2.66 $\pm$ 0.01 & 2.60 $\pm$ 0.01 \\ 
$\Delta\gamma$ & 0.21 $\pm$ 0.02 & 0.28 $\pm$ 0.03 & 0.26 $\pm$ 0.05 
& 0.21 $\pm$ 0.03 & 0.15 $\pm$ 0.05 \\ 
$s$ & 1.70 $\pm$ 0.23 & 2.05 $\pm$ 0.42 & 2.12 $\pm$ 0.84 & 5.86 $\pm$ 2.97 & 5 (fixed) \\
\hline
\end{tabular}
}
\label{Table:fit_para_hard}
\end{table*}

\begin{table*}[htbp]
\centering
\caption{The parameters of the SBPL model obtained from the fit of the spectral softening features for proton, helium, carbon, oxygen and iron. }
\resizebox{\columnwidth}{!}{
\begin{tabular}{c|ccccc}
\hline
\hline
Particle & Proton & Helium & Carbon & Oxygen & Iron \\
\hline
Fit energy range  & 1.58 - 100 TeV & 3.98 - 250 TeV & 19.0 - 675 TeV &  9.0 - 900 TeV & 236 - 1771 TeV  \\ \hline
Nuisance parameters & 3 & 3 & 2 & 3 & 3 \\ 
\hline
$\Phi_0$ ($10^{-5}$ GeV$^{-1}$ m$^{-2}$ s$^{-1}$ sr$^{-1}$) & 8.08 $\pm$ 0.48
& 5.03 $\pm$ 0.35 & 0.60 $\pm$ 0.12 & 1.54 $\pm$ 0.16 & 1.50 $\pm$ 0.47 \\
$E_{\rm br}$ (TeV) & 14.6 $\pm$ 1.3 & 30.6 $\pm$ 4.4  & 96.5 $\pm$ 33.5  
& 123 $\pm$ 38 & 360 $\pm$ 157 \\ 
$\gamma$ & 2.57 $\pm$ 0.01 & 2.40 $\pm$ 0.01 & 2.32 $\pm$ 0.06 
& 2.45 $\pm$ 0.03 & 2.46 $\pm$ 0.05 \\ 
$\Delta\gamma$ & -0.34 $\pm$ 0.05 & -0.32 $\pm$ 0.07 & -0.49 $\pm$ 0.17 
& -0.48 $\pm$ 0.18 & -0.69 $\pm$ 0.41 \\ 
$s$ & 5 (fixed) & 5 (fixed) & 10 (fixed) & 10 (fixed) & 10 (fixed) \\
\hline
\end{tabular}
}
\label{Table:fit_para_soft}
\end{table*}

\noindent{\bf Interpretations}\\
\noindent{\bf (I) The background plus nearby source model.}
This model assumes that the ensemble of many sources contributes to the majority of the 
observed CRs (called as background), with the addition of a nearby one which gives distinct 
features from the background. We adopt the GALPROP propagation code\cite{Strong:1998pw} to 
calculate the fluxes from the background component. The propagation halo is characterized by 
a cylinder with radius $r_h$ and half-height $z_h$. We work in a two-halo propagation 
framework as indicated by the slow diffusion of particles in the vicinities of middle-aged
pulsars\cite{HAWC:2017kbo,LHAASO:2021crt}. To reproduce the observed breaks of the 
secondary-to-primary ratios\cite{DAMPE:2022jgy}, the rigidity-dependence of the diffusion 
coefficient in the Milky Way halo is parameters as a broken PL form, 
$D_{\rm halo}({R})=\beta D_0 (R/{\rm GV})^{\delta_1}[1+(R/R_{\rm br})^2]^{(\delta_2-\delta_1)/2}$,
where $\delta_1$ and $\delta_2$ are the slopes below and above $R_{\rm br}$, $\beta$ is the
particle velocity in unit of light speed. The diffusion coefficients in the disk and halo 
connect with each other smoothly, via $D_{\rm disk}=f(z)D_{\rm halo}$, where
$f(z)=\xi+(1-\xi)[1-\exp(-z^2/2h^2)]$. The propagation parameters are summarized in 
Table \ref{Table:model_para}.

\begin{table*}[htbp]
\caption{\bf Model parameters used to give the results shown in Figure \ref{fig-3}.}
\centering
\begin{tabular}{c|c|c}
\hline
  Parameter & Background + nearby source & Propagation \\
\hline
$r_h$ (kpc) & 20.0 & 20.0 \\
$z_h$ (kpc) & 5.0 & 5.0 \\
$D_0$ ($10^{28}$ cm$^2$~s$^{-1}$) & 2.42 & 2.06 \\
$\delta_1$ & 0.48 & 0.53 \\
$\delta_2$ & 0.20 & 0.13 \\
$R_{\rm br}$ (GV) & 300 & 700 \\
$\delta_3$ & ... & 0.55 \\
$R_{\rm br,2}$ (TV) & ... & 15 \\ 
$\xi$ & 0.1 & 0.1 \\
$h$ (kpc) & 0.45 & 0.45 \\
\hline
$\alpha_{\rm p}$ & 2.48 &  2.36 \\
$\alpha_{\rm he}$ & 2.38 & 2.28 \\
$\alpha_{\rm c}$ & 2.40 & 2.30 \\
$\alpha_{\rm o}$ & 2.42 & 2.32 \\
$\alpha_{\rm fe}$ & 2.44 & 2.34 \\
$N_{\rm p}^{a}$ & $6.60\times10^2$ & $6.98\times10^2$ \\
$N_{\rm he}^{a}$ & $8.91\times10^1$ & $9.65\times10^1$ \\
$N_{\rm c}^{a}$ & 1.07 & 1.12 \\
$N_{\rm o}^{a}$ & 1.00 & 1.00 \\
$N_{\rm fe}^{a}$ & $6.00\times10^{-2}$ & $5.80\times10^{-2}$\\
\hline
$t$ (yr) & $3.4\times10^5$ & ... \\
$r$ (kpc) & 0.25 & ... \\
$\beta$ & 1.90 & ... \\
$R_c$ (TV) & 30 & ... \\
$q_{\rm p}^{b}$ & $1.20\times10^2$ & ... \\
$q_{\rm he}^{b}$ & $3.80\times10^1$ & ... \\
$q_{\rm c}^{b}$ & 1.32 & ... \\
$q_{\rm o}^{b}$ & 1.00 & ... \\
$q_{\rm fe}^{b}$ & 0.14 & ... \\
\hline
Modulation potential (MV) & 650 & 650 \\
\hline
\end{tabular}\\
\label{Table:model_para}
Notes. $^a$Relative (background) source normalization parameters at 1 TV normalized to that of oxygen.
$^{b}$Relative normalizations of the nearby source component at 1 TV normalized to that of oxygen.
\end{table*}

\begin{figure}[!htb]
\begin{center}
\includegraphics[width=0.7\columnwidth]{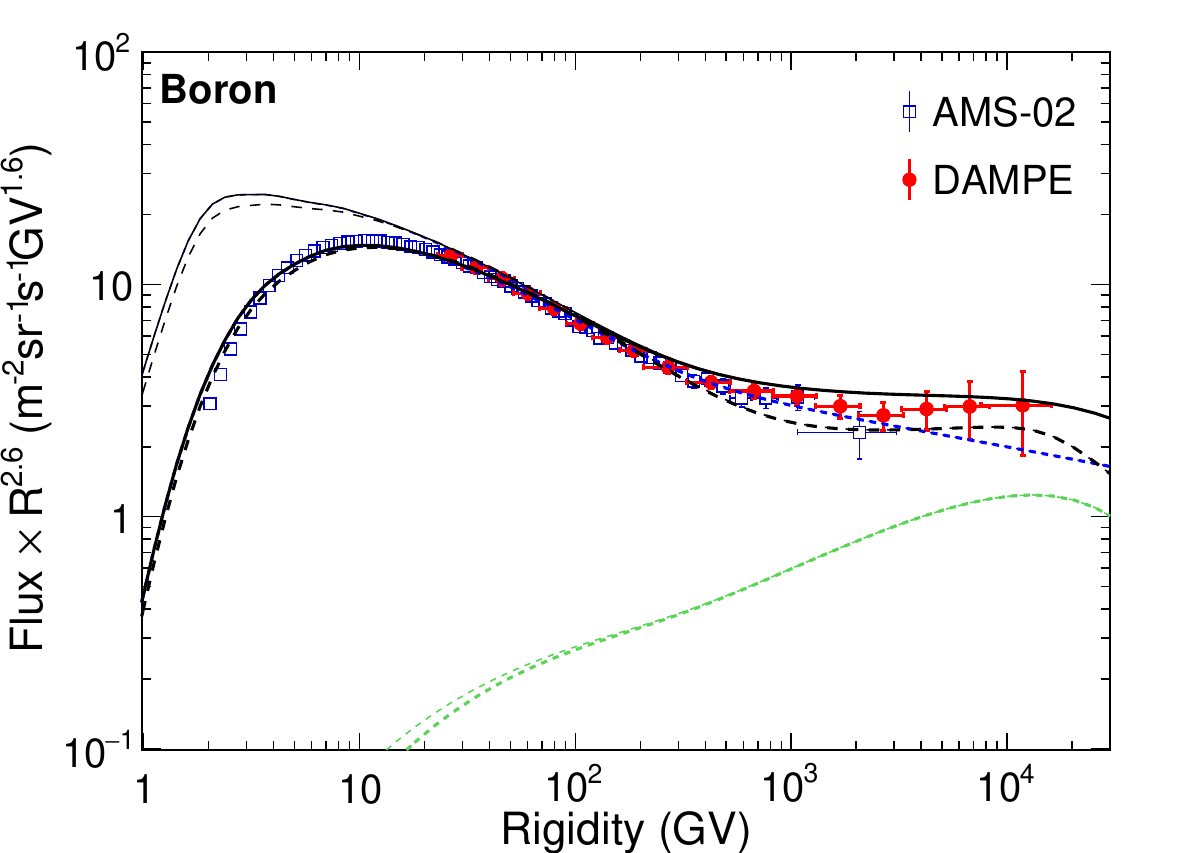}
\end{center}
\caption{{\bf Rigidity spectra of boron weighted by $R^{2.6}$.}
The solid lines show the results from the background plus nearby source model, with individual 
contributions from the two components being shown by dotted lines (blue for the background and 
green for the nearby source). The dashed lines show the results from the propagation model. 
Thin lines correspond to the results in the local interstellar medium and thick lines are 
modulated fluxes with potential of 650 MV. The measurements of AMS-02\cite{AMS:2021nhj} and
DAMPE\cite{DAMPE:2024qwc} are shown for comparison.
}
\label{fig-boron}
\end{figure}

\begin{figure}[!ht]
\begin{center}
\includegraphics[width=0.48\columnwidth]{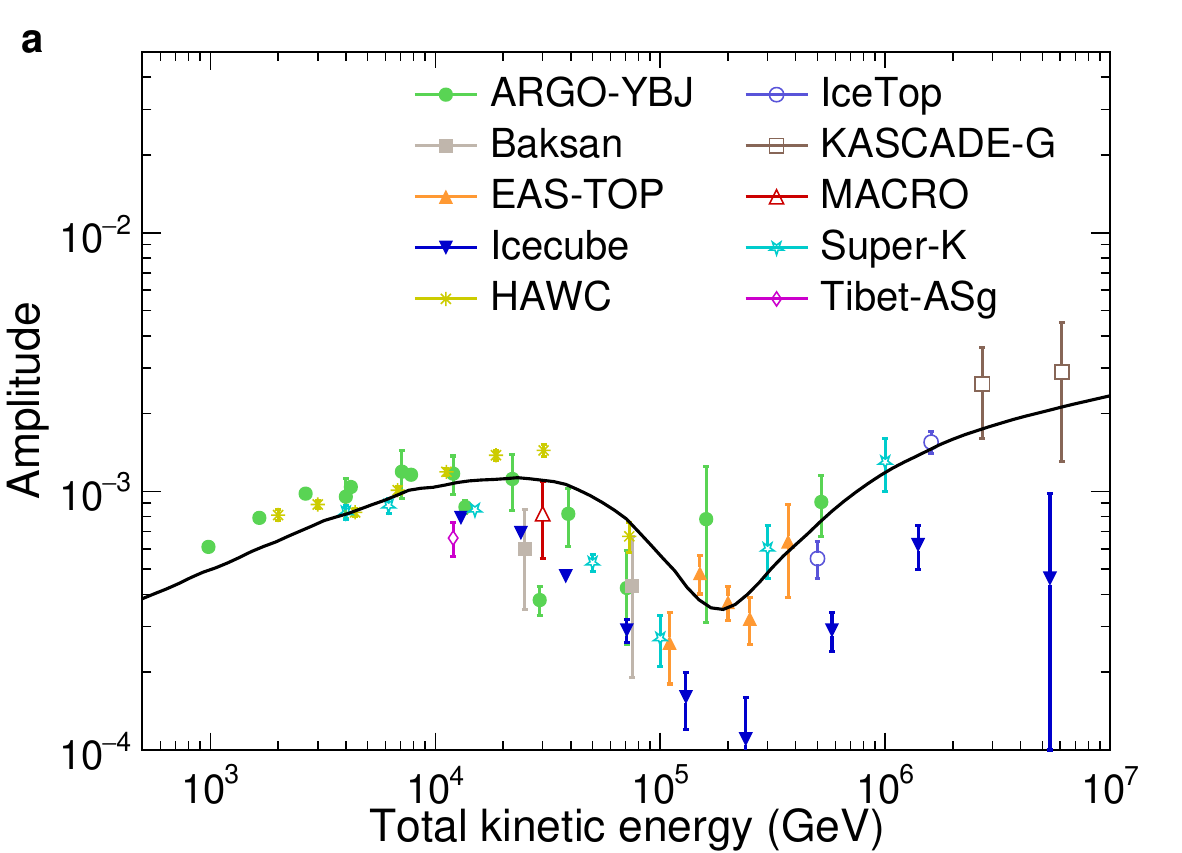}
\includegraphics[width=0.48\columnwidth]{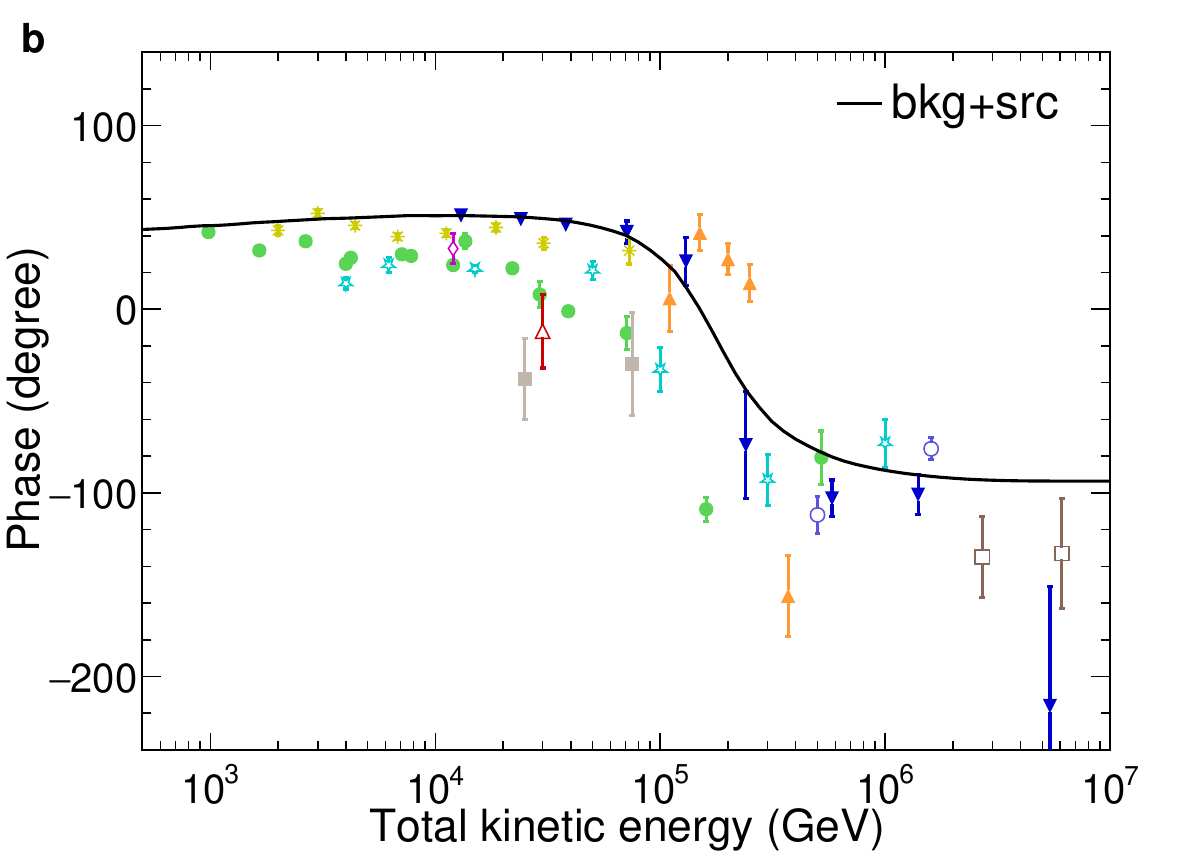}
\end{center}
\caption{{\bf Dipole anisotropies of CRs.}
Panel {\bf a} shows the amplitudes of the dipole anisotropies, and panel {\bf b} shows
the phases. The solid lines show the predictions from the background plus nearby source model.
The measurements are from: 
ARGO-YBJ\cite{ARGO-YBJ:2018zoa}, Baksan\cite{Alekseenko:2009ew}, EAS-TOP\cite{EAS-TOP:2009nld},
HAWC\cite{Abeysekara:2018qho}, IceCube\cite{IceCube:2016biq}, IceTop\cite{IceCube:2016biq},
KASCADE-Grande\cite{2015ICRC...34..281C}, MACRO\cite{MACRO:2002qsl},
Super-K\cite{Super-Kamiokande:2005plj}, Tibet-AS$\gamma$\cite{Amenomori:2017jbv}.}
\label{fig-aniso}
\end{figure}

\begin{figure}[!htb]
\begin{center}
\includegraphics[width=0.7\columnwidth]{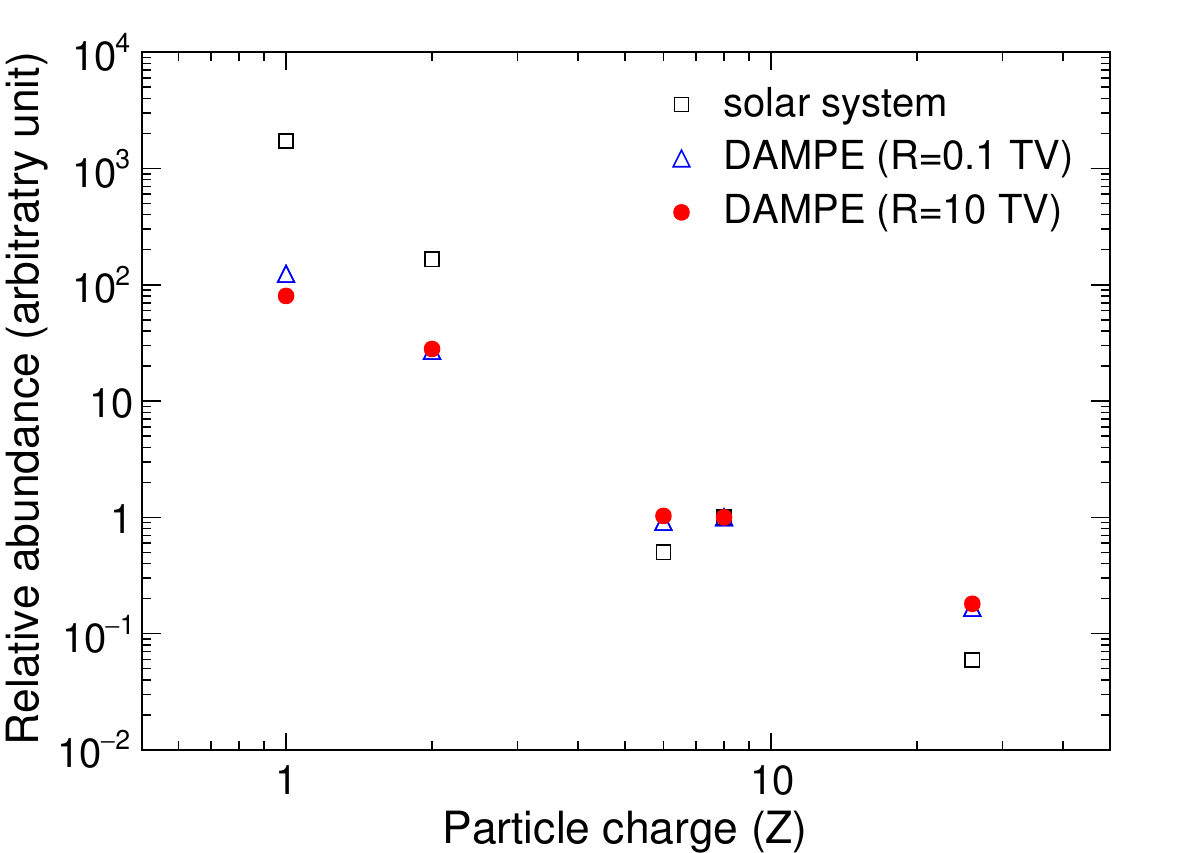}
\end{center}
\caption{{\bf Relative abundances (normalized to oxygen) of the measurements.}
Blue triangles show the results at 0.1 TV, red dots show those at 10 TV, 
and black squares show the solar system abundances\cite{Lodders:2003vvq}. 
}
\label{fig-abun}
\end{figure}

The injection spectrum is assumed to be a PL in rigidity, 
$q_{\rm bkg}(R) \propto (R/{\rm GV})^{-\alpha}$. In the energy range we are interested in, 
the single PL injection is adequate, although more complicated injection form is required to 
reproduce the data in a wider energy range\cite{Boschini:2020jty}. The spectral indices for
different species differ slightly from each other, as given in Table \ref{Table:model_para}.
The spatial distribution is assumed to follow the Galactic SNR distribution\cite{Case:1998qg}, 
but with parameters slightly tuned based on $\gamma$-ray data\cite{Trotta:2010mx}. 
The predicted secondary flux from the background CRs is consistent with the
data\cite{AMS:2021nhj,DAMPE:2024qwc} for rigidity below $\sim$ TV, as can be seen in Figure \ref{fig-boron}. To better compare with the low-energy AMS-02 data, 
a force-field solar modulation with potential of 650 MV has been applied\cite{Gleeson:1968zza}. 
There might also be contribution from secondary interactions occurring around the sources 
given those particles may get trapped for some time\cite{Sun:2023ibg,Yang:2024igs}
(see, however, ref.\cite{Recchia:2021vfw} for a pessimistic opinion).

For the nearby source, we assume an instantaneous injection at time $t$ and location ${\bf r}$.
We assume the nearby source is Geminga-like\cite{Zhao:2021css}, which has $r=250$ pc and 
$t=3.4\times10^5$ yr\cite{Smith:1994xyz,Manchester:2004bp}. Note that the distance and age 
of the nearby source are different from those of ref.\cite{Savchenko:2015dha}, in which a 
$\sim2$ Myr old nearby source with $\sim200$ pc distance was assumed. The main difference 
is that, in ref.\cite{Savchenko:2015dha}, the nearby source dominates the fluxes and 
anisotropies in the $1-100$ TeV band, while in our case, the contribution to the fluxes from 
the nearby source is sub-dominant and the anisotropies are jointly contributed by both components. 
The injection spectrum is assumed to be an exponential cutoff power-law (ECPL) form, 
$q_{\rm src} \propto (R/{\rm GV})^{-\beta}\exp(-R/R_c)$. 
The propagated flux at Earth can be obtained via the Green's function as, 
$\psi(r,t,R)=\frac{q_{\rm src}} {(\sqrt{2\pi}\sigma)^3}\exp\left(-\frac{r^2}{2\sigma^2}\right)$, 
where $\sigma(R,t)=[2D_{\rm disk}(R)t]^{1/2}$ is the effective diffusion length within time $t$.
The slow diffusion coefficient $D_{\rm disk}$ is adopted. We find that for $\beta=1.9$ and 
$R_c=30$ TV, the bump features of spectra can be reproduced, as shown in Figure 3.
Here the proton and oxygen spectra are shown for illustration. The same solar modulation 
potential of 650 MV is adopted.
The expected boron flux from CRs produced by the nearby source is calculated with the
same Green's function method through an integration over the volume and time assuming a gas 
density of 1 hydrogen cm$^{-3}$, which is also shown in Figure \ref{fig-boron}. 
The grammage for CRs from the nearby source can be estimated as $X\sim0.3$ g cm$^{-2}$
for the assumed gas density and a propagation time of $\sim2\times10^5$ yr. We find that the 
contribution from the nearby source is sub-dominant in the whole energy range. Adding such 
a component can nevertheless better reproduce the data at the highest rigidity part.

The dipole component of the large-scale anisotropies from the model can be obtained as
$\Delta=3|\nabla(D_{\rm disk}\psi)|/(c\psi)$. Since the gradients from the background sources
and the nearby source are different, a vector sum is necessary. 
We assume that the Galactic coordinate of the nearby source is $(l,b)=(165^{\circ},-17^{\circ})$, 
which is close to but not exactly at the anti-Galactic-center direction. The location of the 
source in the sky is consistent with the fitting results obtained in ref.\cite{Zhao:2021css}. 
The resulting amplitude and phase (the right ascension of the maximum CR intensity) are shown 
in Figure \ref{fig-aniso}., which reproduce properly the measurements. 
Particularly, at high energies (above a few hundred TeV), the anisotropy is dominated by the 
background which has a phase of $\sim -90^{\circ}$ (Galactic center). At low energies, the
anisitropy is in turn dominated by the nearby source, and a phase flip exists. 
There is also wide discussion about the effect of regulating the propagation of CRs by the local
magnetic field\cite{Battaner:2009zf,Schwadron:2014mho,Mertsch:2014cua}. The anisotropic diffusion in 
the directions perpendicular or parallel to the magnetic field would result in a projection of the 
CR flow onto the direction of the magnetic field\cite{Schwadron:2014mho,Ahlers:2016njd,Liu:2019cun}. 
In such a case, the original direction of the CR source becomes less correlated with the observed 
anisotropy direction, and the constraint on the location of the source is relaxed. In addition, it 
has been shown by the joint observation of HAWC and IceCube that there is a sharp division between 
excesses and deficits of the large-scale anisotropies, and there are also high-order anisotropies 
beyond the dipole\cite{HAWC:2018wju}. More detailed studies are necessary to fully understand these
observational facts of Galactic CRs.

The relative abundances (normalized to ${\rm O}=1$) of the measured CR fluxes at $R=0.1$ 
TV where the background sources dominate and $R=10$ TV where the nearby source contributes a
considerable fraction are shown in \ref{fig-abun}. The solar system 
abundances are also shown for reference\cite{Lodders:2003vvq}. One can see that heavier
nuclei become more abundant at $R=10$ TV than those at $R=0.1$ TV, indicating that the nearby 
source tends to have higher metallicity than the background sources.

\noindent{\bf (II) The propagation model.}
The other scenario is to ascribe the spectral breaks to a propagation effect. Two breaks of the 
diffusion coefficient are required to give the hardenings and subsequent softenings of the spectra.
Physically this may be due to the self-generated turbulence with nonlinear Landau and ion-neutral 
damping of magnetohydrodynamic (MHD) waves by CRs, as proposed in ref.\cite{Chernyshov:2022kxk}. 
Here we empirically approximate the equivalent effect as a change of the rigidity dependence 
of the diffusion coefficient as
\begin{equation}
D({R})=\beta D_0 (R/{\rm GV})^{\delta_1}[1+(R/R_{\rm br})^2]^{(\delta_2-\delta_1)/2}
[1+(R/R_{\rm br,2})^5]^{(\delta_3-\delta_2)/5}.
\end{equation}
Since the softening features are sharper than the hardening features in general, we assume
different smoothness parameters of the breaks. Note that we also assume a two-halo propagation 
model, with the spatial dependence of the diffusion coefficient being the same as above. 
The injection spectrum is again assumed to be a single PL form. The model parameters can be 
found in Table \ref{Table:model_para}. The results for the secondary boron spectrum and primary 
proton and oxygen spectra are shown in Figure \ref{fig-boron} and Figure \ref{fig-3} by dashed lines. 
While the spectral features can also be reproduced by the propagation model, whether the observed
complicated behaviors of the large-scale anisotropies can be explained need further studies.

\renewcommand{\refname}{Methods References}

\begin{mybibliography}{99}

\expandafter\ifx\csname url\endcsname\relax
  \def\url#1{\texttt{#1}}\fi
\expandafter\ifx\csname urlprefix\endcsname\relax\def\urlprefix{URL }\fi
\providecommand{\bibinfo}[2]{#2}
\providecommand{\eprint}[2][]{\url{#2}}

\bibitem{DAMPE:2019lxv}
\bibinfo{author}{Ambrosi, G.} \emph{et~al.}
\newblock \bibinfo{title}{{The on-orbit calibration of DArk Matter Particle
  Explorer}}.
\newblock \emph{\bibinfo{journal}{Astropart. Phys.}}
  \textbf{\bibinfo{volume}{106}}, \bibinfo{pages}{18--34}
  (\bibinfo{year}{2019}).
\newblock \eprint{1907.02173}.

\bibitem{Zhao:2022wfw}
\bibinfo{author}{Zhao, C.} \emph{et~al.}
\newblock \bibinfo{title}{{The study of fluorescence response to energy
  deposition in the BGO calorimeter of DAMPE}}.
\newblock \emph{\bibinfo{journal}{Nucl. Instrum. Meth. A}}
  \textbf{\bibinfo{volume}{1029}}, \bibinfo{pages}{166453}
  (\bibinfo{year}{2022}).

\bibitem{Zang:2025erm}
\bibinfo{author}{Zang, J.} \emph{et~al.}
\newblock \bibinfo{title}{{Determination of the absolute energy scale of the
  DAMPE calorimeter with the geomagnetic rigidity cutoff method}}.
\newblock \emph{\bibinfo{journal}{Astropart. Phys.}}
  \textbf{\bibinfo{volume}{173}}, \bibinfo{pages}{103149}
  (\bibinfo{year}{2025}).
\newblock \eprint{2510.03854}.

\bibitem{GEANT4:2002zbu}
\bibinfo{author}{Agostinelli, S.} \emph{et~al.}
\newblock \bibinfo{title}{{GEANT4 - A Simulation Toolkit}}.
\newblock \emph{\bibinfo{journal}{Nucl. Instrum. Meth. A}}
  \textbf{\bibinfo{volume}{506}}, \bibinfo{pages}{250--303}
  (\bibinfo{year}{2003}).

\bibitem{Tykhonov:2021xlb}
\bibinfo{author}{Tykhonov, A.} \emph{et~al.}
\newblock \bibinfo{title}{{TeV-PeV hadronic simulations with DAMPE}}.
\newblock \emph{\bibinfo{journal}{Proc. Sci.}} \textbf{\bibinfo{volume}{ICRC2019}},
  \bibinfo{pages}{143} (\bibinfo{year}{2021}).

\bibitem{Chen:2023koi}
\bibinfo{author}{Chen, Z.-F.} \emph{et~al.}
\newblock \bibinfo{title}{{BGO quenching effect on spectral measurements of
  cosmic-ray nuclei in DAMPE experiment}}.
\newblock \emph{\bibinfo{journal}{Nucl. Instrum. Meth. A}}
  \textbf{\bibinfo{volume}{1055}}, \bibinfo{pages}{168470}
  (\bibinfo{year}{2023}).
\newblock \eprint{2307.12629}.

\bibitem{Bohlen:2014buj}
\bibinfo{author}{B{\"o}hlen, T.~T.} \emph{et~al.}
\newblock \bibinfo{title}{{The FLUKA Code: Developments and Challenges for High
  Energy and Medical Applications}}.
\newblock \emph{\bibinfo{journal}{Nucl. Data Sheets}}
  \textbf{\bibinfo{volume}{120}}, \bibinfo{pages}{211--214}
  (\bibinfo{year}{2014}).

\bibitem{Smart:2005abc}
\bibinfo{author}{{Smart}, D.~F.} \& \bibinfo{author}{{Shea}, M.~A.}
\newblock \bibinfo{title}{{A review of geomagnetic cutoff rigidities for
  earth-orbiting spacecraft}}.
\newblock \emph{\bibinfo{journal}{Adv. Space Res.}}
  \textbf{\bibinfo{volume}{36}}, \bibinfo{pages}{2012--2020}
  (\bibinfo{year}{2005}).

\bibitem{Tykhonov:2022acr}
\bibinfo{author}{Tykhonov, A.} \emph{et~al.}
\newblock \bibinfo{title}{{A deep learning method for the trajectory
  reconstruction of cosmic rays with the DAMPE mission}}.
\newblock \emph{\bibinfo{journal}{Astropart. Phys.}}
  \textbf{\bibinfo{volume}{146}}, \bibinfo{pages}{102795}
  (\bibinfo{year}{2023}).
\newblock \eprint{2206.04532}.

\bibitem{Dong:2018qof}
\bibinfo{author}{Dong, T.} \emph{et~al.}
\newblock \bibinfo{title}{{Charge measurement of cosmic ray nuclei with the
  plastic scintillator detector of DAMPE}}.
\newblock \emph{\bibinfo{journal}{Astropart. Phys.}}
  \textbf{\bibinfo{volume}{105}}, \bibinfo{pages}{31--36}
  (\bibinfo{year}{2019}).
\newblock \eprint{1810.10784}.

\bibitem{Ma:2018brb}
\bibinfo{author}{Ma, P.-X.} \emph{et~al.}
\newblock \bibinfo{title}{{A method for aligning the plastic scintillator
  detector on DAMPE}}.
\newblock \emph{\bibinfo{journal}{Res. Astron. Astrophys.}}
  \textbf{\bibinfo{volume}{19}}, \bibinfo{pages}{082} (\bibinfo{year}{2019}).
\newblock \eprint{1808.05720}.

\bibitem{Yue:2020hmj}
\bibinfo{author}{Yue, C.} \emph{et~al.}
\newblock \bibinfo{title}{{Correction method for the readout saturation of the
  DAMPE calorimeter}}.
\newblock \emph{\bibinfo{journal}{Nucl. Instrum. Meth. A}}
  \textbf{\bibinfo{volume}{984}}, \bibinfo{pages}{164645}
  (\bibinfo{year}{2020}).
\newblock \eprint{2009.09438}.

\bibitem{Serpolla:2025vrj}
\bibinfo{author}{Serpolla, A.} \emph{et~al.}
\newblock \bibinfo{title}{{Machine-learning correction for the calorimeter
  saturation of cosmic-rays ions with the Dark Matter Particle Explorer:
  towards the PeV scale}}.
\newblock \emph{\bibinfo{journal}{Nucl. Instrum. Meth. A}}
  \textbf{\bibinfo{volume}{1085}}, \bibinfo{pages}{171306}
  (\bibinfo{year}{2026}).
\newblock \eprint{2507.06626}.

\bibitem{Fermi-LAT:2017bpc}
\bibinfo{author}{Abdollahi, S.} \emph{et~al.}
\newblock \bibinfo{title}{{Cosmic-ray electron-positron spectrum from 7 GeV to
  2 TeV with the Fermi Large Area Telescope}}.
\newblock \emph{\bibinfo{journal}{Phys. Rev. D}} \textbf{\bibinfo{volume}{95}},
  \bibinfo{pages}{082007} (\bibinfo{year}{2017}).
\newblock \eprint{1704.07195}.

\bibitem{Strong:1998pw}
\bibinfo{author}{Strong, A.~W.} \& \bibinfo{author}{Moskalenko, I.~V.}
\newblock \bibinfo{title}{{Propagation of cosmic-ray nucleons in the galaxy}}.
\newblock \emph{\bibinfo{journal}{Astrophys. J.}}
  \textbf{\bibinfo{volume}{509}}, \bibinfo{pages}{212--228}
  (\bibinfo{year}{1998}).
\newblock \eprint{astro-ph/9807150}.

\bibitem{HAWC:2017kbo}
\bibinfo{author}{Abeysekara, A.~U.} \emph{et~al.}
\newblock \bibinfo{title}{{Extended gamma-ray sources around pulsars constrain
  the origin of the positron flux at Earth}}.
\newblock \emph{\bibinfo{journal}{Science}} \textbf{\bibinfo{volume}{358}},
  \bibinfo{pages}{911--914} (\bibinfo{year}{2017}).
\newblock \eprint{1711.06223}.

\bibitem{LHAASO:2021crt}
\bibinfo{author}{Aharonian, F.} \emph{et~al.}
\newblock \bibinfo{title}{{Extended Very-High-Energy Gamma-Ray Emission
  Surrounding PSR J0622+3749 Observed by LHAASO-KM2A}}.
\newblock \emph{\bibinfo{journal}{Phys. Rev. Lett.}}
  \textbf{\bibinfo{volume}{126}}, \bibinfo{pages}{241103}
  (\bibinfo{year}{2021}).
\newblock \eprint{2106.09396}.

\bibitem{Boschini:2020jty}
\bibinfo{author}{Boschini, M.~J.} \emph{et~al.}
\newblock \bibinfo{title}{{Inference of the Local Interstellar Spectra of
  Cosmic-Ray Nuclei Z {\ensuremath{\leq}} 28 with the GalProp-HelMod
  Framework}}.
\newblock \emph{\bibinfo{journal}{Astrophys. J. Suppl.}}
  \textbf{\bibinfo{volume}{250}}, \bibinfo{pages}{27} (\bibinfo{year}{2020}).
\newblock \eprint{2006.01337}.

\bibitem{Case:1998qg}
\bibinfo{author}{Case, G.~L.} \& \bibinfo{author}{Bhattacharya, D.}
\newblock \bibinfo{title}{{A new sigma-d relation and its application to the
  galactic supernova remnant distribution}}.
\newblock \emph{\bibinfo{journal}{Astrophys. J.}}
  \textbf{\bibinfo{volume}{504}}, \bibinfo{pages}{761} (\bibinfo{year}{1998}).
\newblock \eprint{astro-ph/9807162}.

\bibitem{Trotta:2010mx}
\bibinfo{author}{Trotta, R.} \emph{et~al.}
\newblock \bibinfo{title}{{Constraints on cosmic-ray propagation models from a
  global Bayesian analysis}}.
\newblock \emph{\bibinfo{journal}{Astrophys. J.}}
  \textbf{\bibinfo{volume}{729}}, \bibinfo{pages}{106} (\bibinfo{year}{2011}).
\newblock \eprint{1011.0037}.

\bibitem{Gleeson:1968zza}
\bibinfo{author}{Gleeson, L.~J.} \& \bibinfo{author}{Axford, W.~I.}
\newblock \bibinfo{title}{{Solar Modulation of Galactic Cosmic Rays}}.
\newblock \emph{\bibinfo{journal}{Astrophys. J.}}
  \textbf{\bibinfo{volume}{154}}, \bibinfo{pages}{1011} (\bibinfo{year}{1968}).

\bibitem{Sun:2023ibg}
\bibinfo{author}{Sun, D.-X.}, \bibinfo{author}{Zhang, P.-P.},
  \bibinfo{author}{Yuan, Q.} \emph{et~al.}
\newblock \bibinfo{title}{{Multimessenger observations support cosmic ray
  interactions surrounding acceleration sources}}.
\newblock \emph{\bibinfo{journal}{Phys. Rev. D}}
  \textbf{\bibinfo{volume}{110}}, \bibinfo{pages}{103039}
  (\bibinfo{year}{2024}).
\newblock \eprint{2307.02372}.

\bibitem{Yang:2024igs}
\bibinfo{author}{Yang, R.} \& \bibinfo{author}{Aharonian, F.}
\newblock \bibinfo{title}{{Confinement of relativistic particles in the
  vicinity of accelerators: A key for understanding the anomalies in secondary
  cosmic rays}}.
\newblock \emph{\bibinfo{journal}{Phys. Rev. D}}
  \textbf{\bibinfo{volume}{111}}, \bibinfo{pages}{083040}
  (\bibinfo{year}{2025}).
\newblock \eprint{2410.22199}.

\bibitem{Recchia:2021vfw}
\bibinfo{author}{Recchia, S.} \emph{et~al.}
\newblock \bibinfo{title}{{Grammage of cosmic rays in the proximity of
  supernova remnants embedded in a partially ionized medium}}.
\newblock \emph{\bibinfo{journal}{Astron. Astrophys.}}
  \textbf{\bibinfo{volume}{660}}, \bibinfo{pages}{A57} (\bibinfo{year}{2022}).
\newblock \eprint{2106.04948}.

\bibitem{Smith:1994xyz}
\bibinfo{author}{{Smith}, V.~V.}, \bibinfo{author}{{Cunha}, K.} \&
  \bibinfo{author}{{Plez}, B.}
\newblock \bibinfo{title}{{Is Geminga a runaway member of the Orion
  association?}}
\newblock \emph{\bibinfo{journal}{\aap}} \textbf{\bibinfo{volume}{281}},
  \bibinfo{pages}{L41--L44} (\bibinfo{year}{1994}).

\bibitem{Manchester:2004bp}
\bibinfo{author}{Manchester, R.~N.}, \bibinfo{author}{Hobbs, G.~B.},
  \bibinfo{author}{Teoh, A.} \& \bibinfo{author}{Hobbs, M.}
\newblock \bibinfo{title}{{The Australia Telescope National Facility pulsar
  catalogue}}.
\newblock \emph{\bibinfo{journal}{Astron. J.}} \textbf{\bibinfo{volume}{129}},
  \bibinfo{pages}{1993} (\bibinfo{year}{2005}).
\newblock \eprint{astro-ph/0412641}.

\bibitem{Battaner:2009zf}
\bibinfo{author}{Battaner, E.}, \bibinfo{author}{Castellano, J.} \&
  \bibinfo{author}{Masip, M.}
\newblock \bibinfo{title}{{Galactic magnetic fields and the large-scale
  anisotropy at MILAGRO}}.
\newblock \emph{\bibinfo{journal}{Astrophys. J. Lett.}}
  \textbf{\bibinfo{volume}{703}}, \bibinfo{pages}{L90--L93}
  (\bibinfo{year}{2009}).
\newblock \eprint{0907.2889}.

\bibitem{Schwadron:2014mho}
\bibinfo{author}{Schwadron, N.~A.} \emph{et~al.}
\newblock \bibinfo{title}{{Global Anisotropies in TeV Cosmic Rays Related to
  the Sun{\textquoteright}s Local Galactic Environment from IBEX}}.
\newblock \emph{\bibinfo{journal}{Science}} \textbf{\bibinfo{volume}{343}},
  \bibinfo{pages}{1245026} (\bibinfo{year}{2014}).

\bibitem{Liu:2019cun}
\bibinfo{author}{Liu, W.}, \bibinfo{author}{Lin, S.-j.}, \bibinfo{author}{Hu,
  H.-b.}, \bibinfo{author}{Guo, Y.-q.} \& \bibinfo{author}{Li, A.-f.}
\newblock \bibinfo{title}{{Two Numerical Methods for the 3D Anisotropic
  Propagation of Galactic Cosmic Rays}}.
\newblock \emph{\bibinfo{journal}{Astrophys. J.}}
  \textbf{\bibinfo{volume}{892}}, \bibinfo{pages}{6} (\bibinfo{year}{2020}).
\newblock \eprint{1909.02908}.

\bibitem{HAWC:2018wju}
\bibinfo{author}{Abeysekara, A.~U.} \emph{et~al.}
\newblock \bibinfo{title}{{All-Sky Measurement of the Anisotropy of Cosmic Rays
  at 10 TeV and Mapping of the Local Interstellar Magnetic Field}}.
\newblock \emph{\bibinfo{journal}{Astrophys. J.}}
  \textbf{\bibinfo{volume}{871}}, \bibinfo{pages}{96} (\bibinfo{year}{2019}).
\newblock \eprint{1812.05682}.

\bibitem{Lodders:2003vvq}
\bibinfo{author}{Lodders, K.}
\newblock \bibinfo{title}{{Solar System Abundances and Condensation
  Temperatures of the Elements}}.
\newblock \emph{\bibinfo{journal}{Astrophys. J.}}
  \textbf{\bibinfo{volume}{591}}, \bibinfo{pages}{1220--1247}
  (\bibinfo{year}{2003}).

\bibitem{ARGO-YBJ:2018zoa}
\bibinfo{author}{Bartoli, B.} \emph{et~al.}
\newblock \bibinfo{title}{{Galactic Cosmic-Ray Anisotropy in the Northern
  Hemisphere from the ARGO-YBJ Experiment during 2008{\textendash}2012}}.
\newblock \emph{\bibinfo{journal}{Astrophys. J.}}
  \textbf{\bibinfo{volume}{861}}, \bibinfo{pages}{93} (\bibinfo{year}{2018}).
\newblock \eprint{1805.08980}.

\bibitem{Alekseenko:2009ew}
\bibinfo{author}{Alekseenko, V.~V.} \emph{et~al.}
\newblock \bibinfo{title}{{10-100 TeV cosmic ray anisotropy measured at Baksan
  EAS 'Carpet' array}}.
\newblock \emph{\bibinfo{journal}{Nucl. Phys. B Proc. Suppl.}}
  \textbf{\bibinfo{volume}{196}}, \bibinfo{pages}{179--182}
  (\bibinfo{year}{2009}).
\newblock \eprint{0902.2967}.

\bibitem{EAS-TOP:2009nld}
\bibinfo{author}{Aglietta, M.} \emph{et~al.}
\newblock \bibinfo{title}{{Evolution of the cosmic ray anisotropy above
  10{\textasciicircum}{14} eV}}.
\newblock \emph{\bibinfo{journal}{Astrophys. J. Lett.}}
  \textbf{\bibinfo{volume}{692}}, \bibinfo{pages}{L130--L133}
  (\bibinfo{year}{2009}).
\newblock \eprint{0901.2740}.

\bibitem{Abeysekara:2018qho}
\bibinfo{author}{Abeysekara, A.~U.} \emph{et~al.}
\newblock \bibinfo{title}{{Observation of Anisotropy of TeV Cosmic Rays with
  Two Years of HAWC}}.
\newblock \emph{\bibinfo{journal}{Astrophys. J.}}
  \textbf{\bibinfo{volume}{865}}, \bibinfo{pages}{57} (\bibinfo{year}{2018}).
\newblock \eprint{1805.01847}.

\bibitem{IceCube:2016biq}
\bibinfo{author}{Aartsen, M.~G.} \emph{et~al.}
\newblock \bibinfo{title}{{Anisotropy in Cosmic-ray Arrival Directions in the
  Southern Hemisphere Based on six Years of Data From the Icecube Detector}}.
\newblock \emph{\bibinfo{journal}{Astrophys. J.}}
  \textbf{\bibinfo{volume}{826}}, \bibinfo{pages}{220} (\bibinfo{year}{2016}).
\newblock \eprint{1603.01227}.

\bibitem{2015ICRC...34..281C}
\bibinfo{author}{{Chiavassa}, A.} \emph{et~al.}
\newblock \bibinfo{title}{{A study of the first harmonic of the large scale
  anisotropies with the KASCADE-Grande experiment}}.
\newblock In \emph{\bibinfo{booktitle}{34th International Cosmic Ray Conference
  (ICRC2015)}}, vol.~\bibinfo{volume}{34} of
  \emph{\bibinfo{series}{Int. Cosmic Ray Conf.}},
  \bibinfo{pages}{281} (\bibinfo{year}{2015}).

\bibitem{MACRO:2002qsl}
\bibinfo{author}{Ambrosio, M.} \emph{et~al.}
\newblock \bibinfo{title}{{The Search for the sidereal and solar diurnal
  modulations in the total MACRO muon data set}}.
\newblock \emph{\bibinfo{journal}{Phys. Rev. D}} \textbf{\bibinfo{volume}{67}},
  \bibinfo{pages}{042002} (\bibinfo{year}{2003}).
\newblock \eprint{astro-ph/0211119}.

\bibitem{Super-Kamiokande:2005plj}
\bibinfo{author}{Guillian, G.} \emph{et~al.}
\newblock \bibinfo{title}{{Observation of the anisotropy of 10-TeV primary
  cosmic ray nuclei flux with the super-kamiokande-I detector}}.
\newblock \emph{\bibinfo{journal}{Phys. Rev. D}} \textbf{\bibinfo{volume}{75}},
  \bibinfo{pages}{062003} (\bibinfo{year}{2007}).
\newblock \eprint{astro-ph/0508468}.

\bibitem{Amenomori:2017jbv}
\bibinfo{author}{Amenomori, M.}
\newblock \bibinfo{title}{{Northern sky Galactic Cosmic Ray anisotropy between
  10-1000 TeV with the Tibet Air Shower Array}}.
\newblock \emph{\bibinfo{journal}{Astrophys. J.}}
  \textbf{\bibinfo{volume}{836}}, \bibinfo{pages}{153} (\bibinfo{year}{2017}).
\newblock \eprint{1701.07144}.

\end{mybibliography}


\begin{thebibliography}{99}

\expandafter\ifx\csname url\endcsname\relax
  \def\url#1{\texttt{#1}}\fi
\expandafter\ifx\csname urlprefix\endcsname\relax\def\urlprefix{URL }\fi
\providecommand{\bibinfo}[2]{#2}
\providecommand{\eprint}[2][]{\url{#2}}

\bibitem{Peters:1961mxb}
\bibinfo{author}{Peters, B.}
\newblock \bibinfo{title}{{Primary cosmic radiation and extensive air
  showers}}.
\newblock \emph{\bibinfo{journal}{Nuovo Cim.}} \textbf{\bibinfo{volume}{22}},
  \bibinfo{pages}{800--819} (\bibinfo{year}{1961}).

\bibitem{Hillas:1984ijl}
\bibinfo{author}{Hillas, A.~M.}
\newblock \bibinfo{title}{{The Origin of Ultrahigh-Energy Cosmic Rays}}.
\newblock \emph{\bibinfo{journal}{Ann. Rev. Astron. Astrophys.}}
  \textbf{\bibinfo{volume}{22}}, \bibinfo{pages}{425--444}
  (\bibinfo{year}{1984}).

\bibitem{Bell:2013vxa}
\bibinfo{author}{Bell, A.~R.}
\newblock \bibinfo{title}{{Cosmic ray acceleration}}.
\newblock \emph{\bibinfo{journal}{Astropart. Phys.}}
  \textbf{\bibinfo{volume}{43}}, \bibinfo{pages}{56--70}
  (\bibinfo{year}{2013}).

\bibitem{Cesarsky:1980pm}
\bibinfo{author}{Cesarsky, C.~J.}
\newblock \bibinfo{title}{{Cosmic Ray Confinement in the Galaxy}}.
\newblock \emph{\bibinfo{journal}{Ann. Rev. Astron. Astrophys.}}
  \textbf{\bibinfo{volume}{18}}, \bibinfo{pages}{289--319}
  (\bibinfo{year}{1980}).

\bibitem{Savchenko:2015dha}
\bibinfo{author}{Savchenko, V.}, \bibinfo{author}{Kachelrie{\ss}, M.} \&
  \bibinfo{author}{Semikoz, D.~V.}
\newblock \bibinfo{title}{{Imprint of a 2 Million Year old Source on the
  Cosmic-ray Anisotropy}}.
\newblock \emph{\bibinfo{journal}{Astrophys. J. Lett.}}
  \textbf{\bibinfo{volume}{809}}, \bibinfo{pages}{L23} (\bibinfo{year}{2015}).
\newblock \eprint{1505.02720}.

\bibitem{Ahlers:2016njd}
\bibinfo{author}{Ahlers, M.}
\newblock \bibinfo{title}{{Deciphering the Dipole Anisotropy of Galactic Cosmic
  Rays}}.
\newblock \emph{\bibinfo{journal}{Phys. Rev. Lett.}}
  \textbf{\bibinfo{volume}{117}}, \bibinfo{pages}{151103}
  (\bibinfo{year}{2016}).
\newblock \eprint{1605.06446}.

\bibitem{Liu:2018fjy}
\bibinfo{author}{Liu, W.}, \bibinfo{author}{Guo, Y.-Q.} \&
  \bibinfo{author}{Yuan, Q.}
\newblock \bibinfo{title}{{Indication of nearby source signatures of cosmic
  rays from energy spectra and anisotropies}}.
\newblock \emph{\bibinfo{journal}{J. Cosmol. Astropart. Phys.}} \textbf{\bibinfo{volume}{10}},
  \bibinfo{pages}{010} (\bibinfo{year}{2019}).
\newblock \eprint{1812.09673}.

\bibitem{Chernyshov:2022kxk}
\bibinfo{author}{Chernyshov, D.~O.}, \bibinfo{author}{Dogiel, V.~A.},
  \bibinfo{author}{Ivlev, A.~V.}, \bibinfo{author}{Erlykin, A.~D.} \&
  \bibinfo{author}{Kiselev, A.~M.}
\newblock \bibinfo{title}{{Formation of the Cosmic-Ray Halo: The Role of
  Nonlinear Landau Damping}}.
\newblock \emph{\bibinfo{journal}{Astrophys. J.}}
  \textbf{\bibinfo{volume}{937}}, \bibinfo{pages}{107} (\bibinfo{year}{2022}).
\newblock \eprint{2209.12302}.

\bibitem{Strong:2007nh}
\bibinfo{author}{Strong, A.~W.}, \bibinfo{author}{Moskalenko, I.~V.} \&
  \bibinfo{author}{Ptuskin, V.~S.}
\newblock \bibinfo{title}{{Cosmic-ray propagation and interactions in the
  Galaxy}}.
\newblock \emph{\bibinfo{journal}{Ann. Rev. Nucl. Part. Sci.}}
  \textbf{\bibinfo{volume}{57}}, \bibinfo{pages}{285--327}
  (\bibinfo{year}{2007}).
\newblock \eprint{astro-ph/0701517}.

\bibitem{PAMELA:2011mvy}
\bibinfo{author}{Adriani, O.} \emph{et~al.}
\newblock \bibinfo{title}{{PAMELA Measurements of Cosmic-ray Proton and Helium
  Spectra}}.
\newblock \emph{\bibinfo{journal}{Science}} \textbf{\bibinfo{volume}{332}},
  \bibinfo{pages}{69--72} (\bibinfo{year}{2011}).
\newblock \eprint{1103.4055}.

\bibitem{Panov:2009iih}
\bibinfo{author}{Panov, A.~D.} \emph{et~al.}
\newblock \bibinfo{title}{{Energy Spectra of Abundant Nuclei of Primary Cosmic
  Rays from the Data of ATIC-2 Experiment: Final Results}}.
\newblock \emph{\bibinfo{journal}{Bull. Russ. Acad. Sci. Phys.}}
  \textbf{\bibinfo{volume}{73}}, \bibinfo{pages}{564--567}
  (\bibinfo{year}{2009}).
\newblock \eprint{1101.3246}.

\bibitem{AMS:2017seo}
\bibinfo{author}{Aguilar, M.} \emph{et~al.}
\newblock \bibinfo{title}{{Observation of the Identical Rigidity Dependence of
  He, C, and O Cosmic Rays at High Rigidities by the Alpha Magnetic
  Spectrometer on the International Space Station}}.
\newblock \emph{\bibinfo{journal}{Phys. Rev. Lett.}}
  \textbf{\bibinfo{volume}{119}}, \bibinfo{pages}{251101}
  (\bibinfo{year}{2017}).

\bibitem{AMS:2021nhj}
\bibinfo{author}{Aguilar, M.} \emph{et~al.}
\newblock \bibinfo{title}{{The Alpha Magnetic Spectrometer (AMS) on the
  international space station: Part II {\textemdash} Results from the first
  seven years}}.
\newblock \emph{\bibinfo{journal}{Phys. Rept.}} \textbf{\bibinfo{volume}{894}},
  \bibinfo{pages}{1--116} (\bibinfo{year}{2021}).

\bibitem{DAMPE:2019gys}
\bibinfo{author}{An, Q.} \emph{et~al.}
\newblock \bibinfo{title}{{Measurement of the cosmic-ray proton spectrum from
  40 GeV to 100 TeV with the DAMPE satellite}}.
\newblock \emph{\bibinfo{journal}{Sci. Adv.}} \textbf{\bibinfo{volume}{5}},
  \bibinfo{pages}{eaax3793} (\bibinfo{year}{2019}).
\newblock \eprint{1909.12860}.

\bibitem{Alemanno:2021gpb}
\bibinfo{author}{Alemanno, F.} \emph{et~al.}
\newblock \bibinfo{title}{{Measurement of the cosmic ray helium energy spectrum
  from 70 GeV to 80 TeV with the DAMPE space mission}}.
\newblock \emph{\bibinfo{journal}{Phys. Rev. Lett.}}
  \textbf{\bibinfo{volume}{126}}, \bibinfo{pages}{201102}
  (\bibinfo{year}{2021}).
\newblock \eprint{2105.09073}.

\bibitem{DAMPE:2022jgy}
\bibinfo{author}{Alemanno, F.} \emph{et~al.}
\newblock \bibinfo{title}{{Detection of spectral hardenings in cosmic-ray
  boron-to-carbon and boron-to-oxygen flux ratios with DAMPE}}.
\newblock \emph{\bibinfo{journal}{Sci. Bull.}} \textbf{\bibinfo{volume}{67}},
  \bibinfo{pages}{2162--2166} (\bibinfo{year}{2022}).
\newblock \eprint{2210.08833}.

\bibitem{DAMPE:2024qwc}
\bibinfo{author}{Alemanno, F.} \emph{et~al.}
\newblock \bibinfo{title}{{Observation of a Spectral Hardening in Cosmic Ray
  Boron Spectrum with the DAMPE Space Mission}}.
\newblock \emph{\bibinfo{journal}{Phys. Rev. Lett.}}
  \textbf{\bibinfo{volume}{134}}, \bibinfo{pages}{191001}
  (\bibinfo{year}{2025}).
\newblock \eprint{2412.11460}.

\bibitem{CALET:2022vro}
\bibinfo{author}{Adriani, O.} \emph{et~al.}
\newblock \bibinfo{title}{{Observation of Spectral Structures in the Flux of
  Cosmic-Ray Protons from 50~GeV to 60~TeV with the Calorimetric Electron
  Telescope on the International Space Station}}.
\newblock \emph{\bibinfo{journal}{Phys. Rev. Lett.}}
  \textbf{\bibinfo{volume}{129}}, \bibinfo{pages}{101102}
  (\bibinfo{year}{2022}).
\newblock \eprint{2209.01302}.

\bibitem{CALET:2023nif}
\bibinfo{author}{Adriani, O.} \emph{et~al.}
\newblock \bibinfo{title}{{Direct Measurement of the Cosmic-Ray Helium Spectrum
  from 40~GeV to 250~TeV with the Calorimetric Electron Telescope on the
  International Space Station}}.
\newblock \emph{\bibinfo{journal}{Phys. Rev. Lett.}}
  \textbf{\bibinfo{volume}{130}}, \bibinfo{pages}{171002}
  (\bibinfo{year}{2023}).
\newblock \eprint{2304.14699}.

\bibitem{Atkin:2018wsp}
\bibinfo{author}{Atkin, E.} \emph{et~al.}
\newblock \bibinfo{title}{{New Universal Cosmic-Ray Knee near a Magnetic
  Rigidity of 10 TV with the NUCLEON Space Observatory}}.
\newblock \emph{\bibinfo{journal}{J. Exp. Theor. Phys. Lett.}} \textbf{\bibinfo{volume}{108}},
  \bibinfo{pages}{5--12} (\bibinfo{year}{2018}).
\newblock \eprint{1805.07119}.

\bibitem{Choi:2022aht}
\bibinfo{author}{Choi, G.~H.} \emph{et~al.}
\newblock \bibinfo{title}{{Measurement of High-energy Cosmic-Ray Proton
  Spectrum from the ISS-CREAM Experiment}}.
\newblock \emph{\bibinfo{journal}{Astrophys. J.}}
  \textbf{\bibinfo{volume}{940}}, \bibinfo{pages}{107} (\bibinfo{year}{2022}).

\bibitem{Yoon:2017qjx}
\bibinfo{author}{Yoon, Y.~S.} \emph{et~al.}
\newblock \bibinfo{title}{{Proton and Helium Spectra from the CREAM-III
  Flight}}.
\newblock \emph{\bibinfo{journal}{Astrophys. J.}}
  \textbf{\bibinfo{volume}{839}}, \bibinfo{pages}{5} (\bibinfo{year}{2017}).
\newblock \eprint{1704.02512}.

\bibitem{Gorbunov:2018stf}
\bibinfo{author}{Grebenyuk, V.} \emph{et~al.}
\newblock \bibinfo{title}{{Energy spectra of abundant cosmic-ray nuclei in the
  NUCLEON experiment}}.
\newblock \emph{\bibinfo{journal}{Adv. Space Res.}}
  \textbf{\bibinfo{volume}{64}}, \bibinfo{pages}{2546--2558}
  (\bibinfo{year}{2019}).
\newblock \eprint{1809.05333}.

\bibitem{KK1958}
\bibinfo{author}{{Kulikov}, G.~V.} \& \bibinfo{author}{{Kristiansen}, G.~B.}
\newblock \bibinfo{title}{{On the Size Spectrum of Extensive Air Showers}}.
\newblock \emph{\bibinfo{journal}{Zh. Eksper. Teor. Fiz.}}
  \textbf{\bibinfo{volume}{35}}, \bibinfo{pages}{635} (\bibinfo{year}{1958}).

\bibitem{Adriani:2020wyg}
\bibinfo{author}{Adriani, O.} \emph{et~al.}
\newblock \bibinfo{title}{{Direct Measurement of the Cosmic-Ray Carbon and
  Oxygen Spectra from 10 GeV$/n$ to 2.2 TeV$/n$ with the Calorimetric Electron
  Telescope on the International Space Station}}.
\newblock \emph{\bibinfo{journal}{Phys. Rev. Lett.}}
  \textbf{\bibinfo{volume}{125}}, \bibinfo{pages}{251102}
  (\bibinfo{year}{2020}).
\newblock \eprint{2012.10319}.

\bibitem{AMS:2021lxc}
\bibinfo{author}{Aguilar, M.} \emph{et~al.}
\newblock \bibinfo{title}{{Properties of Iron Primary Cosmic Rays: Results from
  the Alpha Magnetic Spectrometer}}.
\newblock \emph{\bibinfo{journal}{Phys. Rev. Lett.}}
  \textbf{\bibinfo{volume}{126}}, \bibinfo{pages}{041104}
  (\bibinfo{year}{2021}).

\bibitem{CALET:2021fks}
\bibinfo{author}{Adriani, O.} \emph{et~al.}
\newblock \bibinfo{title}{{Measurement of the Iron Spectrum in Cosmic Rays from
  10 GeV/$n$ to 2.0 TeV/$n$ with the Calorimetric Electron Telescope on the
  International Space Station}}.
\newblock \emph{\bibinfo{journal}{Phys. Rev. Lett.}}
  \textbf{\bibinfo{volume}{126}}, \bibinfo{pages}{241101}
  (\bibinfo{year}{2021}).
\newblock \eprint{2106.08036}.

\bibitem{DAMPE:2017cev}
\bibinfo{author}{Chang, J.} \emph{et~al.}
\newblock \bibinfo{title}{{The DArk Matter Particle Explorer mission}}.
\newblock \emph{\bibinfo{journal}{Astropart. Phys.}}
  \textbf{\bibinfo{volume}{95}}, \bibinfo{pages}{6--24} (\bibinfo{year}{2017}).
\newblock \eprint{1706.08453}.

\bibitem{DAMPE:2017fbg}
\bibinfo{author}{Ambrosi, G.} \emph{et~al.}
\newblock \bibinfo{title}{{Direct detection of a break in the teraelectronvolt
  cosmic-ray spectrum of electrons and positrons}}.
\newblock \emph{\bibinfo{journal}{Nature}} \textbf{\bibinfo{volume}{552}},
  \bibinfo{pages}{63--66} (\bibinfo{year}{2017}).
\newblock \eprint{1711.10981}.

\bibitem{Yu:2017dpa}
\bibinfo{author}{Yu, Y.} \emph{et~al.}
\newblock \bibinfo{title}{{The plastic scintillator detector for DAMPE}}.
\newblock \emph{\bibinfo{journal}{Astropart. Phys.}}
  \textbf{\bibinfo{volume}{94}}, \bibinfo{pages}{1--10} (\bibinfo{year}{2017}).
\newblock \eprint{1703.00098}.

\bibitem{Azzarello:2016trx}
\bibinfo{author}{Azzarello, P.} \emph{et~al.}
\newblock \bibinfo{title}{{The DAMPE silicon{\textendash}tungsten tracker}}.
\newblock \emph{\bibinfo{journal}{Nucl. Instrum. Meth. A}}
  \textbf{\bibinfo{volume}{831}}, \bibinfo{pages}{378--384}
  (\bibinfo{year}{2016}).

\bibitem{Zhang:2016xkz}
\bibinfo{author}{Zhang, Z.} \emph{et~al.}
\newblock \bibinfo{title}{{The calibration and electron energy reconstruction
  of the BGO ECAL of the DAMPE detector}}.
\newblock \emph{\bibinfo{journal}{Nucl. Instrum. Meth. A}}
  \textbf{\bibinfo{volume}{836}}, \bibinfo{pages}{98--104}
  (\bibinfo{year}{2016}).
\newblock \eprint{1602.07015}.

\bibitem{Huang:2020skz}
\bibinfo{author}{Huang, Y.-Y.} \emph{et~al.}
\newblock \bibinfo{title}{{Calibration and performance of the neutron detector
  onboard of the DAMPE mission}}.
\newblock \emph{\bibinfo{journal}{Res. Astron. Astrophys.}}
  \textbf{\bibinfo{volume}{20}}, \bibinfo{pages}{153} (\bibinfo{year}{2020}).
\newblock \eprint{2005.07828}.

\bibitem{DAMPE:2021qet}
\bibinfo{author}{Alemanno, F.} \emph{et~al.}
\newblock \bibinfo{title}{{Observations of Forbush Decreases of Cosmic-Ray
  Electrons and Positrons with the Dark Matter Particle Explorer}}.
\newblock \emph{\bibinfo{journal}{Astrophys. J. Lett.}}
  \textbf{\bibinfo{volume}{920}}, \bibinfo{pages}{L43} (\bibinfo{year}{2021}).
\newblock \eprint{2110.00123}.

\bibitem{DAgostini:1994fjx}
\bibinfo{author}{D'Agostini, G.}
\newblock \bibinfo{title}{{A Multidimensional unfolding method based on Bayes'
  theorem}}.
\newblock \emph{\bibinfo{journal}{Nucl. Instrum. Meth. A}}
  \textbf{\bibinfo{volume}{362}}, \bibinfo{pages}{487--498}
  (\bibinfo{year}{1995}).

\bibitem{Niu:2022vpn}
\bibinfo{author}{Niu, J.-S.} \& \bibinfo{author}{Liu, J.}
\newblock \bibinfo{title}{{Quantitative study of the hardening in the Alpha
  Magnetic Spectrometer nuclei spectra at a few hundred GV}}.
\newblock \emph{\bibinfo{journal}{Front. Astron. Space Sci.}}
  \textbf{\bibinfo{volume}{9}}, \bibinfo{pages}{1044225}
  (\bibinfo{year}{2022}).
\newblock \eprint{2210.08905}.

\bibitem{AMS:2018tbl}
\bibinfo{author}{Aguilar, M.} \emph{et~al.}
\newblock \bibinfo{title}{{Observation of New Properties of Secondary Cosmic
  Rays Lithium, Beryllium, and Boron by the Alpha Magnetic Spectrometer on the
  International Space Station}}.
\newblock \emph{\bibinfo{journal}{Phys. Rev. Lett.}}
  \textbf{\bibinfo{volume}{120}}, \bibinfo{pages}{021101}
  (\bibinfo{year}{2018}).

\bibitem{Blasi:2012yr}
\bibinfo{author}{Blasi, P.}, \bibinfo{author}{Amato, E.} \&
  \bibinfo{author}{Serpico, P.~D.}
\newblock \bibinfo{title}{{Spectral breaks as a signature of cosmic ray induced
  turbulence in the Galaxy}}.
\newblock \emph{\bibinfo{journal}{Phys. Rev. Lett.}}
  \textbf{\bibinfo{volume}{109}}, \bibinfo{pages}{061101}
  (\bibinfo{year}{2012}).
\newblock \eprint{1207.3706}.

\bibitem{Lagage:1983zz}
\bibinfo{author}{Lagage, P.~O.} \& \bibinfo{author}{Cesarsky, C.~J.}
\newblock \bibinfo{title}{{The maximum energy of cosmic rays accelerated by
  supernova shocks}}.
\newblock \emph{\bibinfo{journal}{Astron. Astrophys.}}
  \textbf{\bibinfo{volume}{125}}, \bibinfo{pages}{249--257}
  (\bibinfo{year}{1983}).

\bibitem{Zhao:2021css}
\bibinfo{author}{Zhao, B.} \emph{et~al.}
\newblock \bibinfo{title}{{Geminga SNR: Possible Candidate of the Local
  Cosmic-Ray Factory}}.
\newblock \emph{\bibinfo{journal}{Astrophys. J.}}
  \textbf{\bibinfo{volume}{926}}, \bibinfo{pages}{41} (\bibinfo{year}{2022}).
\newblock \eprint{2104.07321}.

\bibitem{Bowman:2022okd}
\bibinfo{author}{Bowman, D.~P.}, \bibinfo{author}{Scrandis, R.} \&
  \bibinfo{author}{Seo, E.-S.}
\newblock \bibinfo{title}{{Investigating cosmic ray elemental spectra and the
  atmospheric muon neutrino flux}}.
\newblock \emph{\bibinfo{journal}{Adv. Space Res.}}
  \textbf{\bibinfo{volume}{70}}, \bibinfo{pages}{2703--2713}
  (\bibinfo{year}{2022}).

\bibitem{Recchia:2023mmg}
\bibinfo{author}{Recchia, S.} \& \bibinfo{author}{Gabici, S.}
\newblock \bibinfo{title}{{Origin of the spectral features observed in the
  cosmic-ray spectrum}}.
\newblock \emph{\bibinfo{journal}{Astron. Astrophys.}}
  \textbf{\bibinfo{volume}{692}}, \bibinfo{pages}{A20} (\bibinfo{year}{2024}).
\newblock \eprint{2312.11397}.

\bibitem{Malkov:2021gxd}
\bibinfo{author}{Malkov, M.~A.} \& \bibinfo{author}{Moskalenko, I.~V.}
\newblock \bibinfo{title}{{On the Origin of Observed Cosmic-Ray Spectrum Below
  100 TV}}.
\newblock \emph{\bibinfo{journal}{Astrophys. J.}}
  \textbf{\bibinfo{volume}{933}}, \bibinfo{pages}{78} (\bibinfo{year}{2022}).
\newblock \eprint{2105.04630}.

\bibitem{Mertsch:2014cua}
\bibinfo{author}{Mertsch, P.} \& \bibinfo{author}{Funk, S.}
\newblock \bibinfo{title}{{Solution to the cosmic ray anisotropy problem}}.
\newblock \emph{\bibinfo{journal}{Phys. Rev. Lett.}}
  \textbf{\bibinfo{volume}{114}}, \bibinfo{pages}{021101}
  (\bibinfo{year}{2015}).
\newblock \eprint{1408.3630}.

\bibitem{AMS:2024idr}
\bibinfo{author}{Aguilar, M.} \emph{et~al.}
\newblock \bibinfo{title}{{Properties of Cosmic Deuterons Measured by the Alpha
  Magnetic Spectrometer}}.
\newblock \emph{\bibinfo{journal}{Phys. Rev. Lett.}}
  \textbf{\bibinfo{volume}{132}}, \bibinfo{pages}{261001}
  (\bibinfo{year}{2024}).
  

\end{thebibliography}
\end{document}